\documentclass{IEEEtran}
\usepackage{cite}
\usepackage{amsmath,amssymb,amsfonts}
\usepackage{algorithmic}
\usepackage{graphicx}
\usepackage{textcomp}

\usepackage{url}
\usepackage{subfigure}
\usepackage{color}
\usepackage{mathtools}
\usepackage{dsfont} 
\allowdisplaybreaks

\newtheorem{theorem}{Theorem}[section]
\newtheorem{definition}[theorem]{Definition}
\newtheorem{lemma}[theorem]{Lemma}
\newtheorem{proposition}[theorem]{Proposition}
\newtheorem{corollary}[theorem]{Corollary}

\newcommand{\fsquare}{\vrule height6pt width7pt depth1pt}   % filled square
\newcommand{\myendpf}{\hfill\fsquare \\[0.1in]}

\DeclarePairedDelimiter\floor{\lfloor}{\rfloor}

\def \OO {{O}}
\def \oo {{o}}
\newcommand{\pr}{\mathbb{P}}

\newcommand{\N}{\mathbb{N}}

\newcommand{\kk}{\langle K_{n} \rangle}
\newcommand{\K}{ K_{n}}

\newcommand{\limit}{\underset{n \rightarrow \infty}\lim}
 
\newcommand{\comp}{^{\rm c}}

\newcommand{\hh}{\mathbb{H}(n;\mu,K_n)} %Graph
\newcommand{\HH}{\mathbb{I}(n;\pmb{\mu}^r,\pmb{K}^r_n)} %Graph
\newcommand{\HHcmax}{C_{\rm max}(n;\pmb{\mu}^r, \pmb{K}_n^r )} %Graph
\newcommand{\HHcmaxdn}{C_{\rm max}(n;\pmb{\mu}^r, \pmb{K}_n^r,d_n )} %Graph
\newcommand{\hd}{\mathbb{G}(n;\mu,K_n, d_n)} %Graph

\newcommand{\nodes}{\mathcal{N}}
\newcommand{\hhgc}{\mathbb{H}(n;\tilde{\mu},{K}_{r,n})}
\newcommand{\cmax}{C_{\rm max} (n;\mu, K_n)}

\newcommand{\sdcon}{\mathcal{E}_{n,r}(\mu, K_n)}
\newcommand{\sdcondn}{\mathcal{E}_{n,r}(\mu, K_n, d_n)}
\newcommand{\ssdcondn}{\mathcal{E}_{n}(\mu, K_n, d_n; S)}
\newcommand{\zdn}{\mathcal{Z}_{n}(x_n;\mu, K_n, d_n)}

\newcommand{\cmaxdn}{C_{\rm max} (n;\mu, K_n, d_n)}

\def\BibTeX{{\rm B\kern-.05em{\sc i\kern-.025em b}\kern-.08em
    T\kern-.1667em\lower.7ex\hbox{E}\kern-.125emX}}
\begin{document}
%\title{The Giant Component in Inhomogeneous Random K-out Graphs: Existence, Size, and Robustness to Random Node Failures}
\title{Existence and Size of the Giant Component in Inhomogeneous Random K-out Graphs}
\author{Mansi Sood and Osman Ya\u{g}an 
\thanks{
	This work has been supported in part by the
	National Science Foundation through grant CCF \#1617934, David H. Barakat and LaVerne Owen-Barakat Carnegie Institute of Technology Dean's Fellowship, Pennsylvania Infrastructure Technology Alliance (PITA), CyLab@IoT, and Cylab Presidential Fellowship at Carnegie Mellon University.} \thanks{M. Sood and O. Ya\u{g}an are with Department
of Electrical and Computer Engineering and CyLab,
Carnegie Mellon University, Pittsburgh,
PA, 15213 USA. Email:
{\tt\small \{msood@andrew.cmu.edu, oyagan@ece.cmu.edu\}}}\thanks{A preliminary version of this work was presented at the 59$^{\rm th}$ IEEE Conference of Decision and Control (CDC) 2020 \cite{mansi_cdc}.
}
}

\maketitle
\begin{abstract}
	%Existed studies are limited to connectivity properties that require $K_n=\omega(1)$.
 Random K-out graphs are receiving attention as a model to construct sparse yet well-connected topologies in distributed systems including sensor networks, federated learning, and cryptocurrency networks. In response to the growing heterogeneity in emerging real-world networks, where nodes differ in resources and requirements, \emph{inhomogeneous} random K-out graphs were proposed recently. In this model, first, each of the $n$ nodes is classified as type-1 (respectively, type-2) with probability $\mu$ (respectively, $1-\mu)$ independently from the others, where $0<\mu<1$. Next, each type-1 (respectively, type-2) node draws 1 {\em arc} towards a node (respectively, $K_n$ arcs towards $K_n$ distinct nodes) selected uniformly at random. The orientation of the arcs is ignored yielding the inhomogeneous random K-out graph, denoted by $\mathbb{H}(n;\mu,K_n)$. It was recently established that $\mathbb{H}(n;\mu,K_n)$ is connected {\em with high probability} (whp) if and only if $K_n=\omega(1)$. Motivated by practical settings where establishing links is costly and only a bounded choice of $K_n$ is feasible ($K_n = O(1)$), we study the {\em size} of the largest connected {\em sub-network} of $\mathbb{H}(n;\mu,K_n)$. We first show that the trivial condition of $K_n \geq 2$ for all $n$ is sufficient to ensure that $\mathbb{H}(n;\mu,K_n)$ contains a {\em giant} component of size $n-O(1)$ whp. Next, to model settings where nodes can fail or get compromised, we investigate the size of the largest connected sub-network in $\mathbb{H}(n;\mu,K_n)$ when $d_n$ nodes are selected uniformly at random and removed from the network. We show that if $d_n=O(1)$, a giant component of size $n- \OO(1)$ persists for all $K_n \geq 2$ whp. Further, when $d_n=o(n)$ nodes are removed from $\mathbb{H}(n;\mu,K_n)$, the remaining nodes contain a giant component of size  $n(1-o(1))$ whp for all $K_n \geq 2$. We present numerical results to demonstrate the size of the largest connected component when the number of nodes is finite. %A key design question is how to select the parameters $n$, $\mu$, and $K_n$, to ensure that the network exhibits certain desirable properties with high probability.  
\end{abstract}


\section{Introduction}
\subsection{Background}

The field of random graphs lays the mathematical foundations for modeling and analyzing the complex patterns of interconnections typical to real-world networks including communication networks \cite{barabasi2016network}, social networks \cite{newman2002random}, and biological networks\cite{Newman_2002}. A class of random graphs called the {\em random K-out graph} is one of the earliest known models in the literature \cite{FennerFrieze1982,Bollobas}. The random K-out graph comprising $n$ nodes, denoted by $\mathbb{H}(n;K)$, is constructed as follows. Each of the $n$ nodes draws $K$ edges towards $K$ distinct nodes chosen uniformly at random from among all other nodes. The orientation of the edges is then ignored, yielding an {\em undirected} graph. Random K-out graphs achieve connectivity with probability approaching one in the limit of large $n$ for all $K \geq2$ \cite{FennerFrieze1982, Yagan2013Pairwise}, with the probability of connectivity growing as $1-\Theta({1}/{n^{K^2-1}})$ for all $K \geq 2$\cite{mansi_icc}. Consequently, random K-out graphs get connected very {\em easily}, i.e., with far fewer edges ($O(n)$) as compared to classical random graph models including Erd\H{o}s-R\'enyi (ER) random graphs ($O(n \log n)$). Owing to the simplicity of graph construction and their unique connectivity properties, random K-out graphs are receiving increasing attention as a model to construct sparse, yet well-connected topologies in a fully distributed fashion in ad-hoc networks. Random K-out graphs have received renewed interest for designing securely connected sensor networks \cite{Yagan2013Pairwise}, anonymity preserving cryptocurrency networks \cite{FantiDandelion2018}, and fully decentralized learning networks with differential privacy guarantees \cite{2020dprivacy}.

In the context of wireless sensor networks (WSNs), random K-out graphs have been studied \cite{Yagan2013Pairwise,yagan2012modeling,yavuz2017k,yavuz2015toward}
to model the random {\em pairwise} key predistribution scheme \cite{Haowen_2003}. Along with the original key predistribution scheme proposed by Escheanuer and Gligor \cite{Gligor_2002}, the random pairwise scheme is one of the most widely recognized security protocols for sensor networks. The random pairwise scheme is implemented in two phases. In the first phase, each sensor node is paired {\em offline} with $K$ distinct nodes chosen uniformly at random among all other sensor nodes. Next, a \emph{unique} pairwise key is inserted in the memory of each of the paired sensors. After deployment, two sensor nodes can communicate securely only if they have at least one key in common. The random pairwise key predistribution scheme induces a random K-out graph on the set of the participating sensor nodes\cite{Yagan2013Pairwise}, and therefore the connectivity properties of random K-out graphs are of interest in key management schemes for establishing secure WSNs.

In recent years, the rapid proliferation of affordable sensing and mobile devices has led to advances in applications such as remote-sensing, environmental surveillance \cite{wsnapplications}, and decentralized learning \cite{federated2019advances}.  Such emerging applications increasingly rely on integrating {\em heterogeneous} agents that have different capabilities and (security and connectivity) requirements.  Consequently, the analysis of heterogeneous variants of classical random graph models has emerged as an active area of research.  
\cite{eletrebycdc2018,eletrebyIT20, du2007effective,8606999, Rashad/Inhomo, Yagan/Inhomogeneous}, 
\cite{Barabasi_1999, boccaletti2006complex, Lu2008_applications, Wu2007_applications, Yarvis_2005}.  A heterogeneous variant of random K-out graph known as the {\em inhomogeneous random K-out graph} was proposed recently to model heterogeneous networks secured by random {\em pairwise} key predistribution schemes \cite{eletrebyIT20}. 
\begin{figure}
	%\hspace{-.3cm}
	\centering
	\includegraphics[scale=0.13]{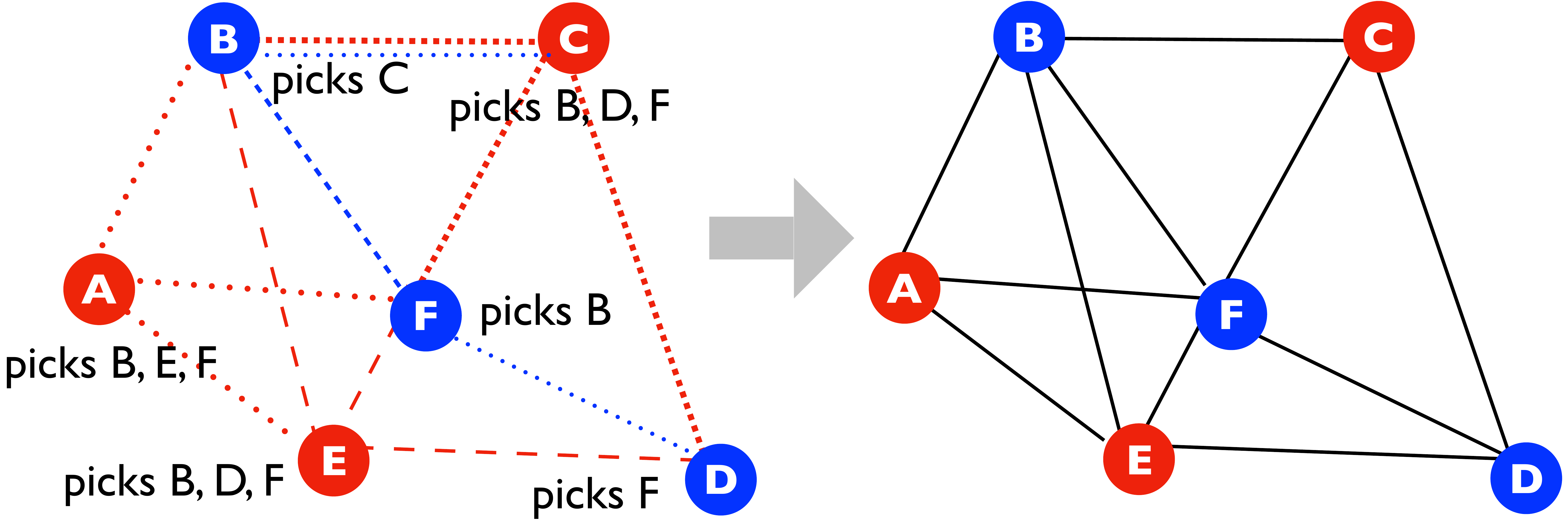} 
	%\vspace{4mm}
	\caption{%Illustration of heterogeneous random pairwise key distribution for
		\sl An inhomogeneous random K-out graph with $6$ nodes. 
		Nodes $A, C$ and $E$ are type-2 and the rest ($B,D,F$) are type-1.
		Each type-1 node selects 1 node uniformly at random and each type-2 node selects $K_n=3$ nodes uniformly at random. An undirected edge is drawn between two nodes if at least one selects the other.}
	\label{fig:pairwise}
\end{figure}

The {\em inhomogeneous} random K-out graph is constructed on a set of $n$ nodes as follows. First, each node is independently classified as type-1 with probability $\mu$ and type-2 with probability $1-\mu$, where $0<\mu<1$. The type of the node determines the number of selections made by it, viz., each type-1 node selects one node uniformly at random from all other nodes and each type-2 node selects $K_n \geq 2$ distinct nodes uniformly at random from all other nodes; see Figure~\ref{fig:pairwise}. The notation $K_n$ indicates that the number of selections made by type-2 nodes {\em scales} as a function of the number of nodes $(n)$. Since \emph{homogeneous }random K-out graphs, where all nodes make the same number of selections, achieve connectivity whp for any $K \geq2$ \cite{FennerFrieze1982, Yagan2013Pairwise}, the interesting case arises when we allow type-1 nodes to select just 1 edge; see Section~\ref{sec:model}. Further, the analysis in this paper also generalizes to inhomogeneous random K-out graphs with an arbitrary number of node types; see Appendix. Throughout our discussion, we let $\hh$ denote the inhomogeneous random K-out graph on $n$ nodes with parameters $\mu$ and $\K$.

% (distributed construction)
%It is known that for $K_1\geq2$ \cite{FennerFrieze1982, Yagan2013Pairwise}, the resulting graph is connected and therefore  the investigation  of connectivity properties of the random K-out graph deal

\subsection{Properties of interest and related results}
{Connectivity} is a fundamental driver of system performance in communication networks since it enables nodes to securely communicate with one another \cite{mao2017connectivity}. 
%
%To provide connectivity guarantees of ad-hoc networks in which the communication links are organized as random K-out graphs
%
%understand how to construct 
%analyze the connectivity of ad-hoc networks generated with random K-out graphs and their heterogeneous variants is to understand the conditions on the network parameters that lead to connectivity.
%A key goal graph construction models such as the random K-out graph i
%Of particular interest to this work is the connectivity of random K-out graphs. 
%with graph construction models such as the random K-out graph
%With distributed graph construction models such as the random K-out graph, a key goal is to understand the conditions on the network paprameters that lead to connectivity.
 %Of particular interest to this work is the connectivity of random K-out graphs. 
%For instance, connectivity enables any pair of nodes to exchange messages in a communication network, or exchange funds in a cryptocurrency network. %
%However, establishing links can be costly and often the goal is to obtain a connected network as efficiently as possible, i.e., by using the least amount of resources (links). 
It is known \cite{Yagan2013Pairwise, FennerFrieze1982} that the random K-out graph $\mathbb{H}(n;K)$ is connected when $K \geq 2$ and \emph{not} connected for $K=1$ with probability tending to one as $n \rightarrow \infty$. In particular, the following zero-one law holds:%
%probability tending to one (as $n \rightarrow \infty$)  %i.e.,  
%if $K \geq 2$, then the resulting graph is 1-connected with high probability. \
%It was established \cite{Yagan2013Pairwise, FennerFrieze1982} that random K-out graphs are connected (respectively, not connected) {\em with high probability} (whp) when $K \geq 2$ (respectively, when $K=1$); i.e., 
\begin{equation} 
	\lim_{n \to \infty} \mathbb{P}\left[ \mathbb{H}(n;K) \text{ is connected}\right] =
	\begin{cases}
		1 & \mathrm{if} \quad K\geq 2, \\
		0 & \mathrm{if} \quad K=1.
	\end{cases}
	\label{eq:homogeneous_zero_one_law}
\end{equation} 
In \cite{eletrebyIT20}, it was shown that for any $\mu \in (0,1)$, 
the \emph{inhomogeneous} random K-out graph $\mathbb{H}(n; \mu,\K)$ is connected with probability tending to one if and only if $K_n=\omega(1)$; i.e.,
%of $\hh$ 
%showed that
\begin{equation} 
	\lim_{n \to \infty} \hspace{-.5mm}\mathbb{P}\left[ \mathbb{H}(n;\mu,K_n) \text{ is connected}\right] \hspace{-.5mm}=\hspace{-.5mm}
	\begin{cases}
		1 & \mathrm{\hspace{-1.5mm}if}~  {\color{black}K_n \rightarrow \infty} \\
		<1 & \mathrm{\hspace{-1.5mm}otherwise}. 
	\end{cases}
	%\hspace{-.5mm}
	\label{eq:introEqcdc}
\end{equation}  
%connected component

As seen from (\ref{eq:introEqcdc}), ensuring connectivity of $\hh$ requires the selections  made by type-2 nodes $(\K)$ to grow unboundedly large with $n$.%which implies an unboundedly large mean node degree\footnote{The mean node degree of $\hh$ is $\frac{2\kk}{n-1}-\left(\frac{\kk}{n-1}\right)^2$; see Appendix $<link>$.}. 
 ~Although it is desirable to have a connected network, in practice, resource constraints can limit the number of links that can be successfully established.  For instance, if the power available for transmission is limited, the underlying physical network may not be dense enough to guarantee global connectivity with key predistribution schemes \cite{hwang2004revisiting}. In such resource-constrained environments where it is infeasible to establish a connected network, the goal is to establish a {\em large connected sub-network} spanning almost the entire network as {\em efficiently} as possible, i.e., by using the least amount of resources (links). \cite{MeiPanconesiRadhakrishnan2008}. For example, if a sensor network is designed to monitor the temperature of a field, it may suffice to aggregate readings from a majority of sensors in the field\cite{liu2006coverage}.
%e Establishing links can be costly and often the goal is to obtain a connected network as {\em efficiently} as possible, i.e., by using the least amount of resources (links).
 %Therefore, the focus of this paper is to investigate conditions on the parameters of $\hh$  provide formal guarantees on  the size of for (component sizes) of $\hh$ when the number of edges drawn by each node is \emph{bounded}, i.e., $K_n=O(1)$.%on the connectivity of  $\hh$ requires $\K=\omega(1)
Formally, the notion of connected sub-networks is described by \emph{connected components}; see Definition \ref{def:concomp} for a precise characterization. With this in mind, the \emph{first} question which we address is when $\K$ is bounded (i.e., $K_n=O(1)$), \emph{how many nodes are contained in the largest connected component of $\hh$?}
%\subsection{Main Contributions}
In the literature on random graphs, this is often studied in terms of the {\em existence} and size of the {\em giant component}, defined as a connected sub-network comprising $\Omega(n)$ nodes; see \cite{erdHos1960evolution} for a classical study on the size of the giant component of Erd\H{o}s-R\'enyi graphs.

Not only is it important to design networks that bear large connected components, but it is also crucial to ensure these large components are \emph{robust} to node \emph{failures}. Sensor networks are often deployed in hostile environments e.g., battlefield monitoring and environmental surveillance, which makes the nodes susceptible to operational failures and adversarial captures. Therefore, it is of interest to analyze the properties of networks induced by  random K-out graphs when some nodes are dishonest, have failed, or have been captured from the network. From a modeling perspective, this translates to analyzing the connectivity properties of inhomogeneous random K-out graphs when a subset of nodes are selected uniformly at random and \emph{deleted}. Based on the analytical framework proposed in this paper, a recent work \cite{elumar2021connectivity} investigates the size of the giant component for \emph{homogeneous} random K-out graphs ($\mathbb{H}(n;K)$) under random node deletions. However, the analysis for $\hh$ has been limited to stronger notions of connectivity that require $\K\rightarrow \infty$ \cite{IT21sood, eletrebyIT20}. With this in mind, the \emph{second} question that we analyze is when $\K$ is bounded, \emph{what is the size of the largest connected component in $\mathbb{H}(n;\mu,K_n)$ after a subset of nodes are selected at random and removed from the network?}
%In this paper, we also analyze the complementary problem of analyzing the {existence of a giant component in $\hh$ when $d_n$ of its nodes are selected uniformly at random and deleted} question?. 

%{\color{blue}[ISIT21 paper- where to cite?]}
%$\K = \omega(1)$ which maybe difficult to achieve in practice.

%depending on the mechanics of node deletions, the following scenariinterest os are of interest: i) when a random number of nodes () are delted   ii) 

%The former has recently studied  and a detailed discussion of $k$-connectivity, i.e., the property that the networks remains connected depite the failure of \emph{any} $k$ nodes, where $k$ is finite. {\color{blue}[ISIT21 paper cite]}
%
%A rel 

\subsection{Main contributions}

%and $\mu \in (0,1)$ 

In this work, we  prove that the inhomogeneous random K-out graph $\hh$ contains a giant component as long as the trivial conditions $\mu \in (0,1)$ and $\K \geq 2$ (for all $n$) hold. In fact, we show that under the same conditions, $\hh$ contains a connected sub-network of size $n- \OO(1)$ whp. Put differently, {\em all but finitely many} nodes are contained in the giant component of $\hh$ whp as $n$ goes to infinity.  This result follows from an upper bound on the probability that more than $M$ nodes are outside of the giant component. We show that this probability decays at least as fast as $O(1)\exp\{-M(1-\mu)(K_n-1)\} +o(1)$ providing a clear trade-off between $K_n$ and the fraction $(1-\mu)$ of nodes that make $K_n$ selections. Next, we generalize this result to study the size of the largest connected component in $\mathbb{H}(n;\mu,K_n)$ after $d_n$ nodes are selected at random and deleted from the network. We show that after the deletion of any randomly chosen \emph{finite} subset of nodes ($d_n=O(1)$), the graph induced by the remaining nodes still contains a giant component of size $n- \OO(1)$ for all $K_n \geq 2$ whp. Furthermore, when $d_n=o(n)$ nodes are deleted uniformly at random from $\mathbb{H}(n;\mu,K_n)$, the remaining nodes contain a giant component of size  $n(1-o(1))$ for all $K_n \geq 2$ whp. We also provide an upper bound on the probability of observing $x_n$ or more nodes outside the giant component in terms of the parameters $d_n, K_n$ and $\mu$. %Our work is the first work that studies the co
For the finite node setting, we provide simulation results investigating the average size and the minimum observed size of the giant component across 100,000 independent experiments. Our simulation results are well-aligned with our theoretical prediction on the existence of a large connected component in the network. For example, with $n=5000, \mu=0.9, \K=2$, at most 45 nodes were observed to be outside the largest connected component across 100,000  experiments; see Section~\ref{sec:Main Results} for details.

\subsection{Further applications of Random K-out Graphs}
Random K-out graphs are a useful tool for constructing distributed networks that are securely connected and resilient to node failures efficiently, i.e., by using the least number of edges. Further, the inhomogeneous random K-out graph allows for more practical deployment scenarios where not all nodes have the same resources or mission requirements. Below, we describe some additional use-cases for random K-out graphs.

There has been a growing interest in {\em fully decentralized} learning applications \cite{federated2019advances} where users communicate over a peer-to-peer network to train a machine learning model. %This mitigates the dependence on a central server for orchestrating the training process. 
In such a decentralized setting, in absence of a trusted aggregator, maintaining the privacy of the local data of participating nodes is a key challenge. A recent line of work \cite{2018dprivacy,2020dprivacy} analyzes mechanisms to perform {\em differentially-private} averaging of private values of users over a communication graph in a fully-decentralized fashion without disclosing the private values. In the proposed GOPA (GOssip Noise for Private Averaging) protocol \cite[Algorithm~1]{2020dprivacy}, each node pair in the communication graph $(u,v)$ collaboratively generates a \emph{pairwise} noise term $\Delta(u,v)$. %Next, one of the nodes adds $\Delta(u,v)$ to their private value while the other node subtracts $\Delta(u,v)$. %Finally, all users add an independent noise term and broadcast the sum of three terms:  
%The pairwise noise term allows the users to mask their their private values without affecting the utility, since the pairwise noise terms get cancelled upon aggregating updates from the users; see \cite{2020dprivacy}. 
%The privacy guarantees \cite[Theorem~1]{2020dprivacy} of GOPA \cite[Algorithm~1]{2020dprivacy} hold as long as the communication graph is connected and the pairwise noise variances are large enough. Furthermore, graphs with better connectivity require a smaller variance for the pairwise noise to attain the desired privacy guarantees. However, the addition of more communication links entails higher communication costs. 
The authors of \cite{2018dprivacy,2020dprivacy} use the random K-out graph as a practical randomized procedure to construct communication graphs that are connected (and thus require a lower pairwise noise variance) and sparse (and thus incur lower communication costs) in a distributed fashion. {A precise characterization of the differential privacy guarantees achieved while using random K-out graphs to construct the communication graph can be found in \cite[Theorem~3]{2020dprivacy}. } The achieved differential-privacy guarantees depend on the connectivity of the subgraph of \emph{honest} users. Further, in cases where the subgraph on honest nodes is not connected, the performance of the proposed protocol depends on the size of connected components of honest users.

In addition to the applications in WSNs and distributed learning, a structure similar to the random K-out graph was proposed \cite[Algorithm~1]{FantiDandelion2018} in the {\em Dandelion} framework for generating anonymity graphs to facilitate the  diffusion of transaction information to protect cryptocurrency networks from mass {\em de-anonymization} attacks. Random K-out graphs have also received interest in the ongoing development of the next-generation SCION Internet architecture to provide end hosts more control over their traffic using path-aware routing\cite{SCION2017book}. In this context, gossip networks closely resembling random K-out graphs have been used to develop distributed systems to efficiently translate between old IP addresses and new SCION addresses \cite{SIAM2021paper}.
%{An ongoing line of research is developing the next-generation SCION Internet architecture to provide end hosts more control over their traffic usinng path-aware routing. Recently, gossip networks closely resembling random K-out graphs have been used to develop distributed systems to translate between old (IP) and new (SCION) addressesthe SCION  Our gossip networks (like the Bitcoin network) closely fit the model of “K-out graphs”, "}
Given their ability to get connected with a relatively smaller number of edges than classical models such as ER graphs, random K-out graphs offer a promising potential for informing the topological properties of emerging peer-to-peer networks. For instance, random K-out graphs may be of interest in Payment Channel Networks \cite{lightning2016} where there is the trade-off between the number of edges in the network (which is constrained since each edge corresponds to funds escrowed in the PCN) and connectivity (which is desirable to facilitate  transactions between participating nodes). Our results on the impact of heterogeneity on the size of the largest connected component of random K-out graphs are of interest for these additional applications.

\subsection{Notation}
We use the following notational conventions throughout the paper. All limits are understood with the number of nodes $n$ going to infinity. While comparing asymptotic behavior of a pair of sequences $\{a_n\},\{b_n\}$, we use $a_n = \oo(b_n)$, $a_n=\omega(b_n)$,  $a_n = \OO(b_n)$, $a_n=\Theta(b_n)$, {\color{black}and $a_n = \Omega(b_n)$} with their meaning in the standard Landau notation. 
All random variables are defined on the same probability triple $(\Omega, {\mathcal{F}}, \mathbb{P})$.
Probabilistic statements are made with respect to this probability measure $\mathbb{P}$, and we denote the corresponding expectation operator by $\mathbb{E}$. 
%We let $\ii\{A\}$ denote the indicator random variable which takes the value 1 if event $A$ occurs and 0 otherwise. 
For an event $A$, its complement is denoted by $A\comp$. %We let $\ii[A]$ denote the indicator random variable which takes the value 1 if event $A$ occurs and 0 otherwise. 
 %For events $A$ and $B$, we let $A\cap B$ denote the intersection of events $A$ and $B$.  
 We say that an event occurs with high probability (whp) 
 %or asymptotically almost surely (a.a.s.) 
 if it holds with \emph{probability tending to one} as $n\rightarrow \infty$.
 %We say that an event occurs with high probability (whp) if it holds with \emph{probability one} as $n\rightarrow \infty$. Sometimes, this is stated equivalently as an event taking place asymptotically almost surely (a.a.s.). 
 We denote the cardinality of a discrete set $A$ by $|A|$ and the set of all positive integers by $\N_0$. For events $A$ and $B$,  we use $A \implies B$ with the meaning that  $A \subseteq B$.

\subsection{Organization}
In Section~\ref{sec:model} we formally define the inhomogeneous random K-out graphs and a model for random node deletions. In Section~\ref{sec:Main Results} we present our main results on characterizing the size of the largest connected component in $\hh$. Here, we provide a discussion of the main results and present simulations for the finite node regime. We provide a proof outline for the existence of a giant component spanning $n-O(1)$ nodes in Section~\ref{sec:proof}, presenting its comprehensive proof in Section~\ref{sec:proof_prop}. Similarly, we outline the proof for the case of node deletions in Section~\ref{sec:proof-deleted} and provide full details in Sections~\ref{sec:proof_prop_deleted}. Lastly, we extend our results to the inhomogeneous random K-out graphs with an arbitrary number of node types in the Appendix.
 \section{Network Model}
\label{sec:model}
\subsection{The Inhomogeneous Random K-out Graph $\hh$}
\label{sec:model-a}
%Consider a network comprising of $n$ nodes {\color{black}indexed by labels} $i=1,2,\dots n$.
Let $\nodes:=\{v_1,v_2,\dots,v_n\}$ denote the set of nodes. The inhomogeneous random K-out graph is constructed on the node set $\nodes$ as follows. First, each node is assigned as type-1 (respectively, type-2) with probability $\mu$ (respectively, $1-\mu$) independently from other nodes, where $0<\mu<1$. 
Next, each type-1 (respectively, type-2) node selects $K_1$ (respectively, $K_2$) distinct nodes uniformly at random among all other nodes. For each $v_i \in \nodes$, let $\Gamma_{n}(i) \subseteq \nodes \setminus v_i$ denote the subset of nodes selected by $v_i$.  Under the aforementioned assumptions,  $\Gamma_{n}(v_1), \dots , \Gamma_{n}(v_n)$ are mutually independent {\em given} the types of nodes. 
We say that  distinct nodes $v_i$ and $v_j$ are adjacent, denoted by $v_i \sim v_j$ if at least one of them picks the other. Namely, 
\begin{align}
%\vspace{-2mm}
v_i \sim v_j ~~\quad \mbox{if} ~~~\quad v_j \in \Gamma_{n}(v_i) ~\vee~ v_i \in \Gamma_{n}(v_j). 
\label{eq:Adjacency}
\end{align}
%Thus, the adjacency condition (\ref{eq:Adjacency}) gives a precise construction of edges on the nodes $\{v_1,\ldots,v_n\}$. 
The inhomogeneous random K-out graph is defined on the node set $\nodes$ through the adjacency condition (\ref{eq:Adjacency}). More general constructions with an arbitrary number of node types are also possible \cite{eletrebyIT20}, and the implications of our results for such cases are presented in the Appendix.

Without loss of generality, we assume that $1\leq K_1<K_2$. From (\ref{eq:homogeneous_zero_one_law}), it can be seen that the inhomogeneous random K-out graph will be connected whp if $K_1 \geq 2$. Therefore, interesting cases arise for the analysis of the connected largest component only when $K_1 = 1$; i.e., when each node has a positive probability $\mu$ of selecting only one other node. As in  \cite{eletrebyIT20}, we thus assume that $K_1=1$ which in turn implies that $K_2 \geq 2$. For generality, we allow $K_2$ to scale with (i.e., to be a function of) $n$  and simplify the notation by denoting the corresponding mapping as $K_n$.  Put differently, we consider the inhomogeneous random K-out graph, denoted as $\hh$, where each of the $n$ nodes selects one other node with probability $\mu$ $(0<\mu<1)$ and $K_n$ other nodes with probability $1-\mu$; the edges are then constructed according to (\ref{eq:Adjacency}). Throughout, we assume that $K_n \geq 2$ for all $n$ in line with the assumption that $K_2 > K_1 =1$. We denote the average number of selections made by each node in $\hh$ by $\kk$. Observe that
\begin{align}
 \kk =   \mu+(1-\mu)K_n.
 \label{eq:avg_K}
\end{align} 
\subsection{Modeling random node failures in $\hh$}
Next, we describe a way to model nodes that are failed and/or compromised, rendering them removed from the resulting network. Given a set of nodes $\nodes:=\{v_1, \ldots, v_n\}$, we first construct an inhomogeneous random K-out graph $\hh$ with parameters $\mu$ and $\K$ as outlined above in Section~\ref{sec:model-a}. The graph $\hh$ constitutes the initial network before any nodes fail or get compromised. Next, we model a random attack on the network which results in the failure/capture of $d_n$ out of $n$ nodes. A subset $D$ of the node set $\nodes$ of size $d_n$ is selected uniformly at random ($D \subset \nodes$ and $|D|=d_n$) from all possible subsets of $\nodes$ containing exactly $d_n$ nodes. We define $\hd$ as the subgraph of $\hh$ induced by the subset of nodes contained in $\nodes \setminus D$. A pair of nodes $(u_1,u_2)$ contained in $\hd$ are adjacent if and only if $(u_1,u_2)$ are adjacent in $\hh$, and $u_1,u_2 \notin D$. The graph $\hd$ represents the resulting sub-network of reliable/honest nodes obtained after removing the failed/compromised nodes in $\hh$.

% Let $d_n$ denote the total number of nodes that are failed and/or compromised. 

%\newpage
%\vspace{15pt}

%such that $|D|=d_n$) (
%
%
%, 
%
%and parameters $\mu$
%
%we construct a random K
%
%
%
%
%
%
%First, we construct an inhomogeneous
%
%Let $D \subset \{v_1, v_2, \dots, v_n \}$ denote the subset of nodes that are removed from the node set $\{v_1, v_2, \dots, v_n \}$ and define $d_n=|D|$. 

%\\ Selections-$\hh$-delection-$\hd$
%{\color{red}TO DO: Add formal model for node deletions + figure}

\section{Main Results and Discussion}
\label{sec:Main Results}

\subsection{Main results}
It is known \cite{eletrebyIT20} that $\hh$ is connected whp {\em only if} $K_n = \omega(1)$. 
%that $\hh$ has a positive probability of being not connected if $K_n = $ . 
A natural question is then to ask what would happen if $\K$ is {\em bounded}, i.e, when $\K=\OO{(1)}$.
It was shown \cite{eletrebyIT20} that $\hh$ has a positive probability of being {\em not} connected in that case. 
Thus, it is of interest to analyze whether the network has a connected sub-network containing a {\em large} number of nodes, or  it consists merely of {\em small} sub-networks isolated from each other. To answer this question, we formally define connected components and then state our main result characterizing the size of the largest connected component of $\hh$ when $K_n = O(1)$.

\begin{definition}[Connected Components]
Nodes $v_1$ and $v_2 \in \nodes$ are said to be {\em connected} if there exists a path of edges connecting them. The connectivity of a pair of nodes forms an equivalence relation on the set of nodes. Consequently, there is a partition of the set of nodes $\nodes$ into non-empty sets  $C_1, C_2, \ldots, C_m$ (referred to as connected components) such that two vertices $v_1$ and $v_2$ are connected if and only if there exists $ i \in \{1, \ldots, m\}$ for which $v_1, v_2 \in C_i$; see \cite[p.~13]{bondy1976graph}.
\label{def:concomp}
\end{definition}{}

\par In light of the above definition, a graph is connected if it consists of only one connected component.  In all other cases, the graph is {\em not} connected and has at least two connected components that have no edges in between. It is of interest to analyze the fraction of the nodes contained in the {\em largest} connected component as the number of nodes grows. In particular, 
a graph  with $n$ nodes is said to have a {\em giant} component if its largest connected component is of size  $\Omega(n)$.

Let $\cmax$ denote the set of nodes in the largest connected component of $\hh$. %Further, we let $\cmaxdn$ denote the largest connected component for $\hh$. %{ \color{blue} [Add formal model description] include details such as deletions after selections}
Our first main result, presented below, shows that $|\cmax| = n - O(1)$ whp. Namely,  
$\hh$ has a giant component that contains {\em all but finitely many} of the nodes whp. 
First, we show that the probability of at least $M$ nodes being {\em outside} of $\cmax$ decays exponentially fast with $M$. 
%\begin{theorem}
%{\sl 
%For the inhomogeneous random graph $\hh$ with $K_n \geq 2 ~\forall n$ and  
%$K_n =O(1)$, for each $M = 1, 2, \ldots$, such that $M \leq n/3$, it holds that 
%\begin{align}
%&  \pr \left[ |\cmax| \leq n-  M\right] \nonumber \\
% &\leq \frac{\exp\{-M\left(\kk-1\right)(1-\oo(1))\}}{1-{\exp\{-\left(\kk-1\right)(1-\oo(1))\}}} + \oo(1) \label{eq:gc_theorem}
%\end{align}
%}  \label{theorem:gc}
%\end{theorem}
\begin{theorem}
	{\sl 
		For the inhomogeneous random graph $\hh$ with $K_n \geq 2 ~\forall n$ and  
		$K_n =O(1)$ for each $M = 1, 2, \ldots$, it holds that 
		\begin{align}
			&  \pr \left[ |\cmax| \leq n-  M\right] \nonumber \\
			&\leq \frac{\exp\{-M\left(\kk-1\right)(1-\oo(1))\}}{1-{\exp\{-\left(\kk-1\right)(1-\oo(1))\}}} + \oo(1) \label{eq:gc_theorem}
		\end{align}
	}  \label{theorem:gc}
\end{theorem}
The proof of Theorem~\ref{theorem:gc} relies on %showing the improbability of existence of \emph{cuts} of size in the range $[M, n-M]$ as described in Section~\ref{sec:proof}. 
the connection between the {\em non-existence} of sub-graphs with size exceeding $M$ that are {\em isolated} from the rest of the graph, and the size of the of largest component being at least $n-M$.
This approach is inspired by \cite{MeiPanconesiRadhakrishnan2008} and differs from the branching process technique typically employed in the random graph literature, e.g., in the case of Erd\H{o}s-R\'enyi graphs \cite[Ch.~4]{van2016random}. The proof of Theorem~\ref{theorem:gc} is presented in Section~\ref{sec:proof}. The next corollary uses Theorem~\ref{theorem:gc} to show that $\hh$ has a giant component that contains {all but finitely many} of the nodes whp.
\begin{corollary}
{\sl For the inhomogeneous K-out random graph $\hh$ with $K_n \geq 2 ~\forall n$ and  
$K_n =O(1)$ we have
\begin{align}{}
|\cmax| = n - O(1) \ \ {\rm whp}.
\label{eq:gc_xn_BigO}
\end{align}
}
\label{cor:gc_2_classes}
\end{corollary}
%\vspace{-5mm}
%\myproof 
The proof of Corollary~\ref{cor:gc_2_classes} is presented in Section~\ref{subsec:cor1}. We extend Corollary~\ref{cor:gc_2_classes} to inhomogeneous random $K$-out graphs with {\em arbitrary} number of node types; see Corollary~\ref{cor:gc_r_classes} in the Appendix.%~\ref{sec:appendix-r-classes}.

So far, we have discussed that $\hh$ contains a giant component of size $n-O(1)$ whp. The next question which we investigate is whether there exists a giant component, i.e., a component of size $\Omega(n)$ in the event of node failures or adversarial capture of nodes in the network. Recall that $\hd$ denotes the random graph obtained by deleting $d_n$ nodes selected at random from $\hh$. Let $\cmaxdn$ denote the set of nodes in the largest connected component of $\hd$. Our next result characterizes the existence and size of the giant component in $\hd$ obtained by deleting $d_n=o(n)$ nodes uniformly at random from $\hh$.%, maybe a figure}%, i.e., the inhomogeneous random K-out graph with random node deletions.
%{\color{red}*update*}
{
\begin{theorem}
	{\sl Consider the graph $\hd$ with \\$d_n=o(n)$, $K_n \geq 2 ~\forall n$ and $K_n =O(1)$. For any sequence $x:\N_0 \rightarrow \N_0$ and $\forall~\epsilon >0$, if $x_n > \frac{(1+\epsilon)d_n}{\kk -1 }~\forall n$, then
\begin{align}
	&  \pr \left[ |\cmaxdn| \leq n-  d_n-x_n\right] \nonumber \\
	&\leq \frac{e^{- x_n (\kk-1) \frac{\epsilon}{1+\epsilon}(1-\oo(1))}}{1-{e^{-\left(\kk-1\right)\frac{\epsilon}{1+\epsilon}(1-\oo(1))}}} + \frac{e^{-{x_n}\left(1-\mu\right)\frac{\epsilon}{1+\epsilon}(1-\oo(1))}}{1-{e^{-\left(1-\mu\right)\frac{\epsilon}{1+\epsilon}(1-\oo(1))}}}.	
%	\frac{\exp\{-x_n\left(\kk-1\right)(1-\oo(1))\}}{1-{\exp\{-\left(\kk-1\right)(1-\oo(1))\}}} + \oo(1). 
\label{eq:gc_general_theorem}
\end{align}}
	\label{theorem:gc-deleted}
\end{theorem}}
Theorem~\ref{theorem:gc-deleted} provides an upper bound on the probability that more than $x_n$ nodes lie outside the largest component $\cmaxdn$ which decays to zero whenever $x_n=\omega(1)$ and $x_n > \frac{(1+\epsilon)d_n}{(\kk -1 )}~\forall n$. 
%Note that for $x_n=\omega(1)$, and given $0<\epsilon_1< \epsilon_2$, 
% a smaller value of $\epsilon$ corresponds to a smaller condition $x_n$ 
Here, the constant $\epsilon$ provides a way to trade-off between the lower bound imposed on $x_n$  $(x_n> {(1+\epsilon)d_n}/{(\kk -1) })$ and the coefficient  ${\epsilon}/{(1+\epsilon)}$ that governs the decay rate of the exponent in the upper bound \eqref{eq:gc_general_theorem} on $\pr \left[ |\cmaxdn| \leq n-  d_n-x_n\right]$; see Section~\ref{sec:proof-deleted} for more details.% In particular, for $x_n=\omega(1)$, if we require 
%{\color{blue}choice of epsilon} %\footnote{Can Theorem~\ref{theorem:gc-deleted} yield Theorem~\ref{theorem:gc}? {\color{blue}check}
	%Note that Theorem~\ref{theorem:gc-deleted} subsumes Theorem~\ref{theorem:gc} in the sense that Theorem~\ref{theorem:gc} can be derived using Theorem~\ref{theorem:gc-deleted} by letting $d_n \rightarrow 0$. 
%	However, the proof of Theorem~\ref{theorem:gc} itself constitutes a crucial first step for proving Theorem~\ref{theorem:gc-deleted}, and through intermediate results (e.g., Lemma~\ref{lem:gc_sum}) provides a general framework for proving the giant component size for random graph models beyond the random K-out graph. Moreover, presenting the analysis for the case without node deletions followed by the case of node deletions allows for a more natural progression of ideas and clearer exposition of the key analytical steps.}. 
The following corollary to Theorem~\ref{theorem:gc-deleted} provides a succinct characterization of the size of $\cmaxdn$.
%	and to allow a more natural progression of ideas, i.e., going from the analysis without node failures 
%	
%	
%	
%	borrows some crucial ideas used 
%
%
%
%%which gives the trivial condition $x_n >0$.% that is satisfied for each $M = 1, 2, dots$ . %  is a stronger result than Theorem~\ref{theorem:gc} 
%However, the proof of Theorem~\ref{gc-deleted} 
%
%for a natural progression of ideas and easier exposition
%
%for a more natural progression the proof 
%subsumes Theorem~\ref{theorem:gc-deleted} but intricate and also empolys some key strategies, for easier exposition, apriori, more natural progression of ideas framework useful
\begin{corollary}
{\sl	For the graph $\hd$ with $K_n \geq 2 ~\forall n$ and $K_n =O(1)$, it holds that
	\begin{enumerate}
		\item[i)] if $d_n=O(1)$, then \begin{align}|\cmaxdn| = n- O(1) \ {\rm whp},\label{eq:gc-deleted-O1}\end{align}
		\item[ii)] if $d_n= \omega(1)$ and $d_n=o(n)$, then \begin{align}|\cmaxdn| = n(1-o(1)) \ {\rm whp}.\label{eq:gc-deleted-on}\end{align}
	\end{enumerate}
%	{\sl For $d_n=O(1)$}
%	\begin{align*}
%		|\cmaxdn| 
%		= n- O(1) \ {\rm whp}.
%	\end{align*}
%	{\sl  For $d_n=o(n)$ }
%\begin{align*}
%	|\cmaxdn| = n(1-o(1)) \ {\rm whp}.
%\end{align*}
\label{cor:gc-deleted}}
\end{corollary}

%\begin{corollary}
%	{\sl For $d_n=O(1)$}
%	\begin{align*}
%		|\cmaxdn| 
%		= n- O(1) \ {\rm whp}.
%	\end{align*}
%	
%	\label{cor:gc-deleted-O1}
%\end{corollary}
%\begin{corollary}
%	{\sl  For $d_n=o(n)$ }
%	\begin{align*}
%		|\cmaxdn| = n(1-o(1)) \ {\rm whp}.
%	\end{align*}
%	
%	\label{cor:gc-deleted-on}
%\end{corollary}
A consequence of Corollary~\ref{cor:gc-deleted} is that despite the deletion of a randomly chosen \emph{finite} subset of nodes ($d_n=O(1)$) in $\hh$, a giant component of size $n- \OO(1)$ persists for all $K_n \geq 2$ whp. Furthermore, when $d_n=o(n)$ nodes are randomly deleted from $\mathbb{H}(n;\mu,K_n)$, the remaining nodes contain a giant component of size $n(1-o(1))$ for all $K_n \geq 2$ whp. The proofs for Theorem~\ref{theorem:gc-deleted} and Corollary~\ref{cor:gc-deleted} are presented in Section~\ref{sec:proof-deleted}.
% \footnote{Define $d_n, x_n$ as scalings?}

\subsection{Discussion}

Theorem~\ref{theorem:gc} shows that for all $\K \geq 2$ and $\mu\in (0,1)$, the largest connected component in $\hh$ spans $n-\OO(1)$ nodes whp. In contrast to the requirement of $\K=\omega(1)$ for connectivity \cite{eletrebyIT20}, merely setting $\K = 2$ ensures that only finitely many nodes $\hh$ lie outside the giant component $\cmax$. We expect that in resource-constrained environments (e.g., IoT type settings), it will be advantageous to have a {\em large} connected component reinforcing the usefulness of inhomogeneous random K-out graphs as a topology design tool in distributed systems. Moreover, for settings where a finite number of nodes fail and/or get captured, modeled by random node deletions, Theorem~\ref{theorem:gc-deleted} implies that the subset of honest nodes still spans $n-\OO(1)$ nodes whp. Furthermore, during large-scale failures and attacks on the network where as many as $d_n=\omega(1), d_n=o(n)$ (e.g., $d_n= n^{0.99}$) nodes are removed from the network, the remaining nodes contain a giant component of size $n(1-o(1))$ whp.

%ensuring 
%Theorem~\ref{theorem:gc-deleted}
%random vs targeted

%the heterogeneous pairwise key predistribution scheme for ensuring secure communications in such applications; see \cite{Haowen_2003,eletrebyIT20} for other advantages of the (heterogeneous) pairwise scheme. 

It is worth emphasizing that the  largest connected component of $\hh$, whose size is given in  (\ref{eq:gc_xn_BigO}), is {\em much larger} than what is strictly required to qualify it as a {\em giant} component, i.e, the condition that $|\cmax| = \Omega(n)$. In fact, for most random graph models, including
Erd\H{o}s-R\'enyi graphs \cite{erdHos1960evolution} and
random key graphs
\cite[Theorem 2]{Rybarczyk_2011}, 
studies on the size of the largest connected component are focused on characterizing the behavior of $|C_{\rm max}|/n$ as $n$ gets large; this amounts to studying the {\em fractional} size of the largest connected component. Our results (\ref{eq:gc_xn_BigO}) and (\ref{eq:gc-deleted-O1}) go beyond looking at the fractional size of the largest component, for which they yield
$ \frac{|\cmax|}{n} \to_{p} 1$. 
%, ~~ {\rm i.e.,} ~~ \nonumber\\
%& |\cmax| = n -o(n) ~~{\rm whp}. 
%\end{align*}{}
This is equivalent to having $|\cmax| = n -o(n)$.
However, even having $|\cmax| = n-\oo(n)$ leaves the possibility that as many as $n^{0.99}$ nodes are {\em not} part of the largest connected component. Thus, our result in Theorem~\ref{theorem:gc}, showing that at most $O(1)$ nodes are outside the largest connected component {\em whp}, is sharper than existing results on the {\em fractional} size of the largest connected component.  Further, even when $d_n=o(n)$ are randomly deleted from $\hh$, \eqref{eq:gc-deleted-on} shows that the fractional size of the giant component of $\hd$ is one whp.

%Our focus is on practical deployment scenarios, where the number of selections made by each node and consequently the mean node degree is \emph{finite}. {\color{blue}first to e.} 
%
%go beyond the requirement of $\K=\omega(1)$
%we need to go beyond 

Our results highlight a major difference of inhomogeneous random K-out graphs with classical models such as Erd\H{o}s-R\'enyi (ER) graphs \cite{Bollobas,erdHos1960evolution}. We provide an example to compare the size of the giant component in $\hh$ and ER graphs with the same mean degree. For $\hh$, we set $\K=2$ and $\mu=0.9$, which yields a mean node degree of $ (1-\oo(1)) 2 \kk  \approx 2 (0.9+0.1\times2)=2.2$; {see Appendix~\ref{sec:mean-node-degree}}. Let $\mathbb{G}(n;p_n)$ denote the ER graph on $n$ nodes with edge probability $p_n \in [0,1]$. We set $p_n=2.2/n$ to get a mean degree of 2.2. Thus, the mean number of edges in both these models {match}. It is known \cite{erdHos1960evolution} that for $p_n=c/n$ and $c>1$, the ER graph has a {giant} component of size $\beta n (1+\oo(1))$ whp, where $\beta \in (0,1]$ is the solution of $\beta + e^{-\beta c}=1$. Substituting $p=2.2/n$, the largest connected component of the ER graph $\mathbb{G}(n;2.2/n)$ is of size $\approx 0.8437n +o(n)$ whp. For an ER graph over $5000$ nodes, this corresponds to over 700 nodes being isolated from the largest component. In contrast, Theorem \ref{theorem:gc} shows that the largest connected component of $\hh$ would be much larger. Namely, $\cmax=n-O(1)$ whp. This is verified in our experiments in Figure~\ref{fig:gc_varymu}, where  for a network of $5000$ nodes, with $\mu \leq 0.9$, at most 45 nodes are outside the largest connected component  in 100,000 experiments.

A related property of interest is the connectivity of $\hh$ under \emph{targeted} attacks when \emph{any} $k-1$ nodes are deleted, where $k \geq 2$.  For $\hh$, it was shown\cite{IT21sood} that \emph{$k$-connectivity}, i.e., the property that the network remains connected despite the failure of any $k-1$ nodes or edges requires $K_n=\Omega(\log n)$ where $k\geq 2$. The requirement for $k$-connectivity ($K_n=\Omega(\log n)$) \cite{IT21sood} is significantly larger than what is required for connectivity ($\K=\omega(1)$) \cite{eletrebyIT20}. These existing studies \cite{eletrebyIT20,IT21sood} focus on stronger notions of connectivity of $\hh$ that require $\K=\omega(1)$. Therefore, our analysis is the first to guarantee a useful property, viz., giant component when the number of selections is finite ($\K=O(1)$), paving the way for practical deployment of inhomogeneous random graphs as a topology design tool in real-world networks.

 \subsection{Simulation results}

 \begin{figure}[!t]
 	\centering
 	\includegraphics[scale=0.31]{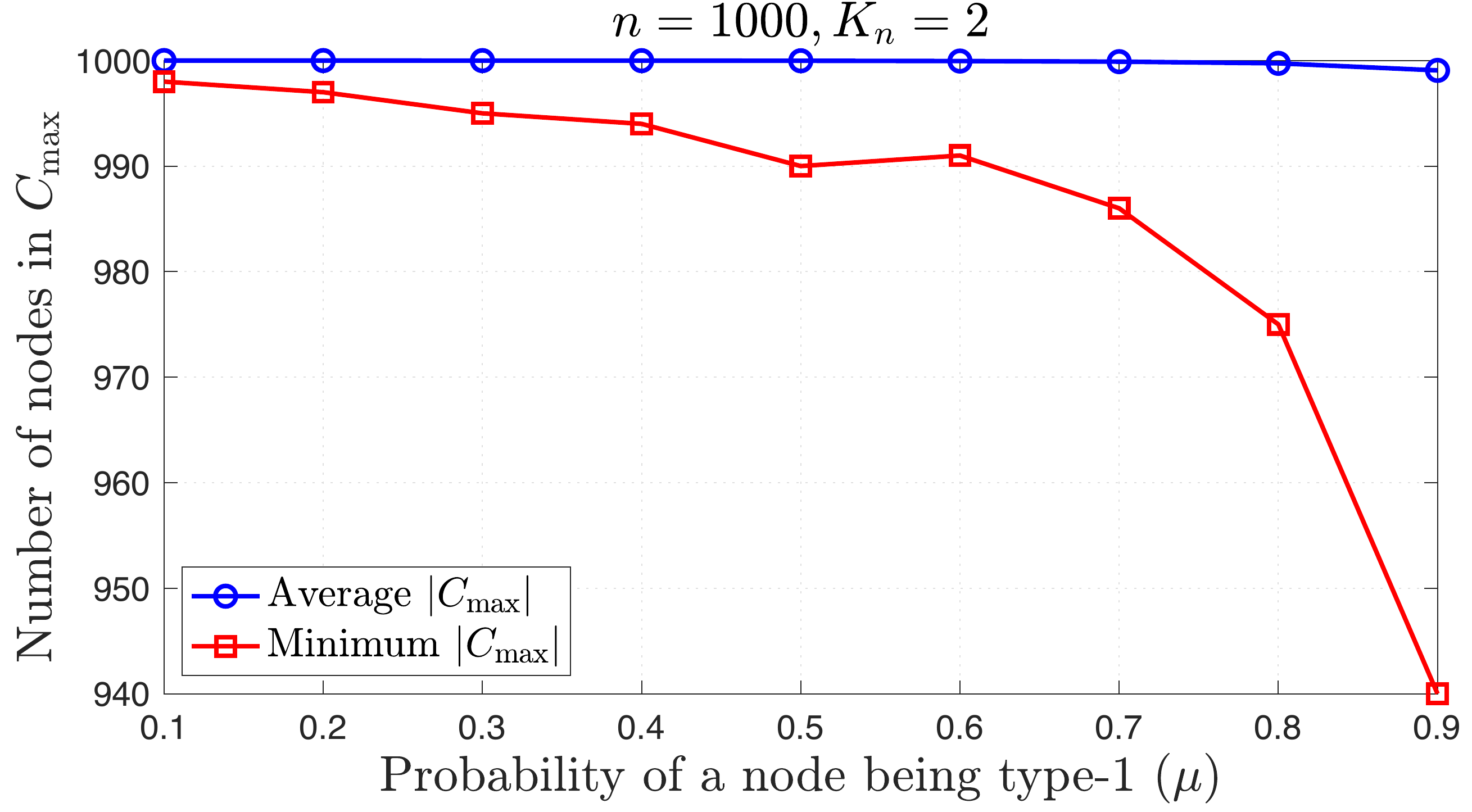}\label{fig:gc1000}\vspace{5mm}
 	\includegraphics[scale=0.31]{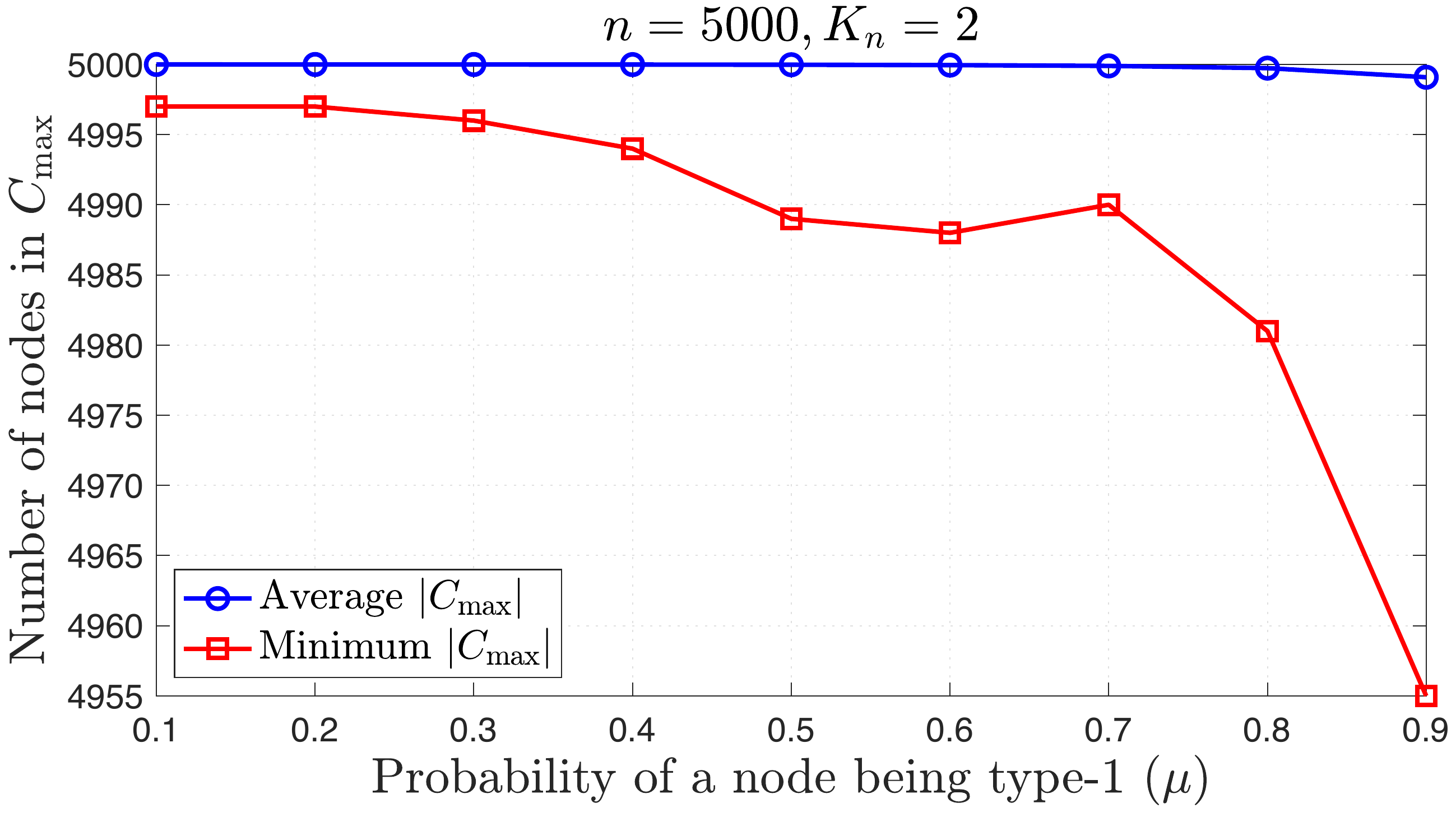}\label{fig:gc5000}
 	\caption{\sl Average  and minimum number of nodes contained in the largest connected component of $~\hh$  with $K_n=2$, $n=1000,5000$ and $\mu \in \{0.1, \dots, 0.9\}$. Even when $\mu=0.9$, setting $K_n=2$ is enough to ensure that almost all of the nodes form a connected component; at most 45 out of 5000 nodes (or, 60 out of 1000 nodes) are seen to be isolated from the giant component across 100,000 independent experiments. } 
 	\label{fig:gc_varymu}
 \end{figure}
 Through simulations, we examine the size of $\cmax$ and $\cmaxdn$ when the number of nodes is finite. We first explore the impact of varying the probability $\mu$ of a node being type-1 on the size of the largest connected component. We generate 100,000 independent realizations of $\hh$ with $\K=2$ for $n=1000$ and  $n=5000$, varying $\mu$ between 0.1 and 0.9 in increments of 0.1. 
 Since, Theorem~\ref{theorem:gc} states that the size of the largest connected component is $n-\OO(1)$ whp, we focus on the {\em minimum} size of the largest component observed in 100,000 experiments. The {\em average} size of the largest connected component is also shown for comparison in  Figure~\ref{fig:gc_varymu}.  
 We see that even when the probability of a node being type-1
 is as high as 0.9, setting $\K=2$ suffices to have a connected component spanning almost all of the nodes. For $n=1000$ and $5000$, at most 60 and 45 nodes, respectively, are found to be outside the largest connected component (Figure~\ref{fig:gc_varymu}). The observation that the number of nodes outside the largest connected component does not scale with $n$ is consistent with Theorem~\ref{theorem:gc} and Corollary~\ref{cor:gc_2_classes}.

   \begin{figure}
  	%\hspace{-.3cm}
  	\centering
  	\includegraphics[scale=0.34]{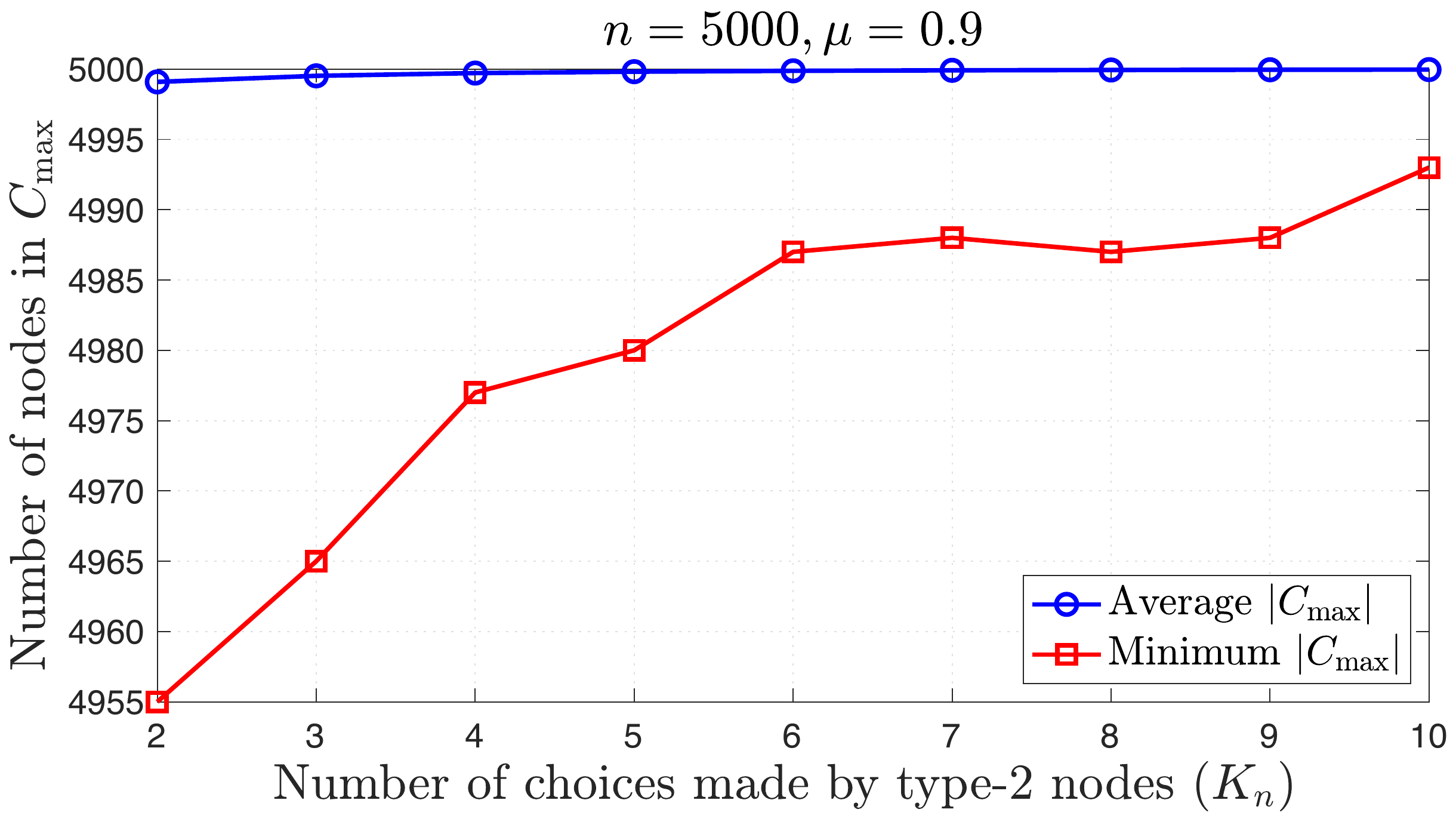} 
  	\caption{\sl Average  and minimum number of nodes contained in the largest connected component of $~\hh$ across 100,000 independent experiments with $n=5000, \mu=0.9$, and $K_n \in \{2, \dots, 10\}$. }
  	\label{fig:gc_varyK}
  	%\vspace{-3m}
  \end{figure}
 The next set of experiments  probes the impact of varying the number $\K$ of edges pushed by type-2 nodes while $\mu$ is fixed. We generate 100,000 independent realizations of $\hh$ for  $n=5000$ while keeping $\mu$ fixed at $0.9$ and varying $\K$ between 2 and 10 in increments of 1. Increasing $\K$ has an impact similar to decreasing $\mu$ and we see in Figure~\ref{fig:gc_varyK} that both the average and the minimum size of the largest connected component increases nearly monotonically. Given that increasing $\K$ (or, decreasing $\mu$) increases $\kk$ in view of
 (\ref{eq:avg_K}), this observation is consistent with
 our main result given in Theorem \ref{theorem:gc}; i.e., with the fact that $\pr \left[   n-|\cmax| > M  \right]$ decays to zero exponentially with $(\kk - 1)M$. 
 
\begin{figure}[ht] 
	\centering
	\includegraphics[scale=0.34]{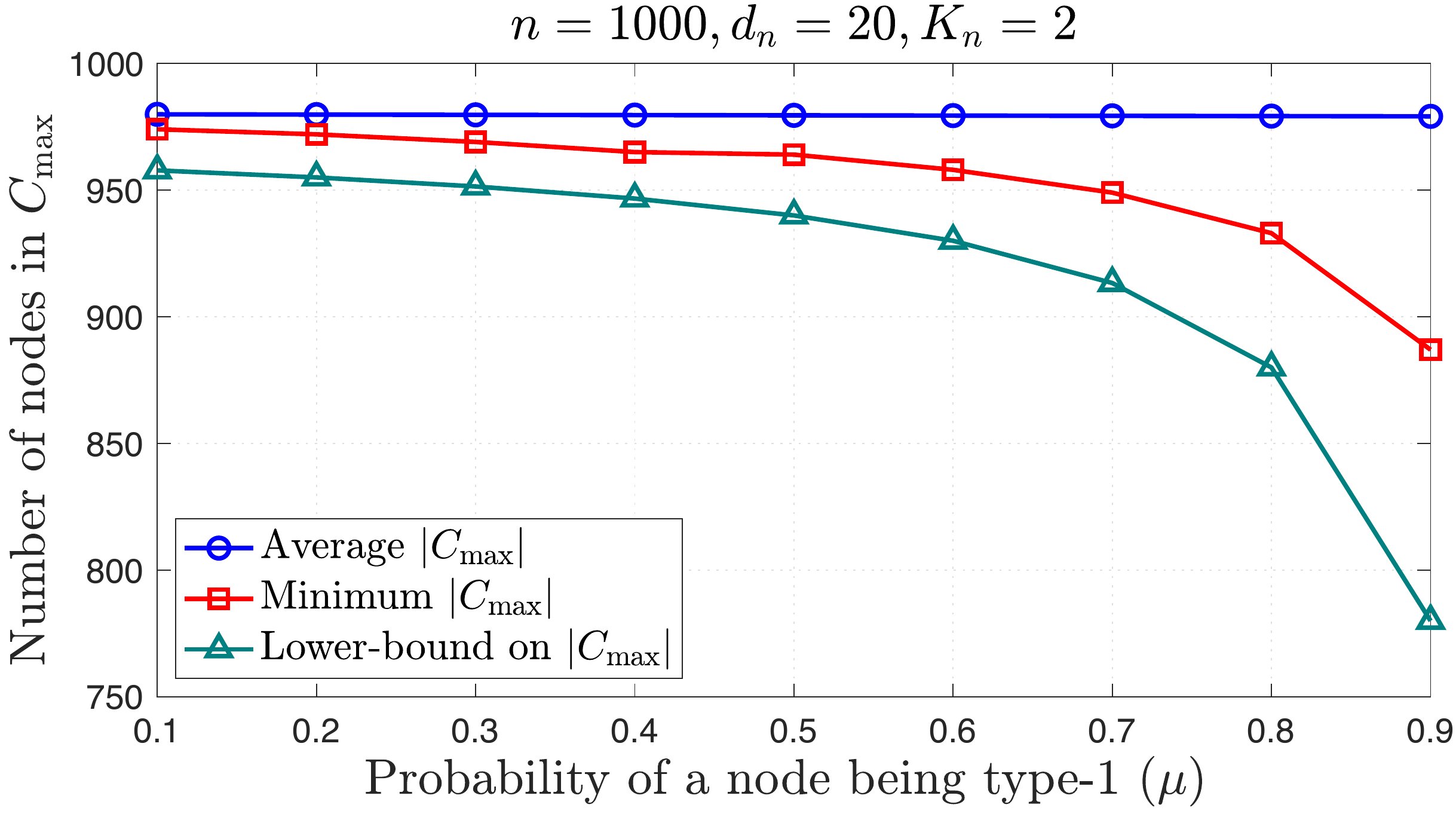}\vspace{5mm}
	\includegraphics[scale=0.34]{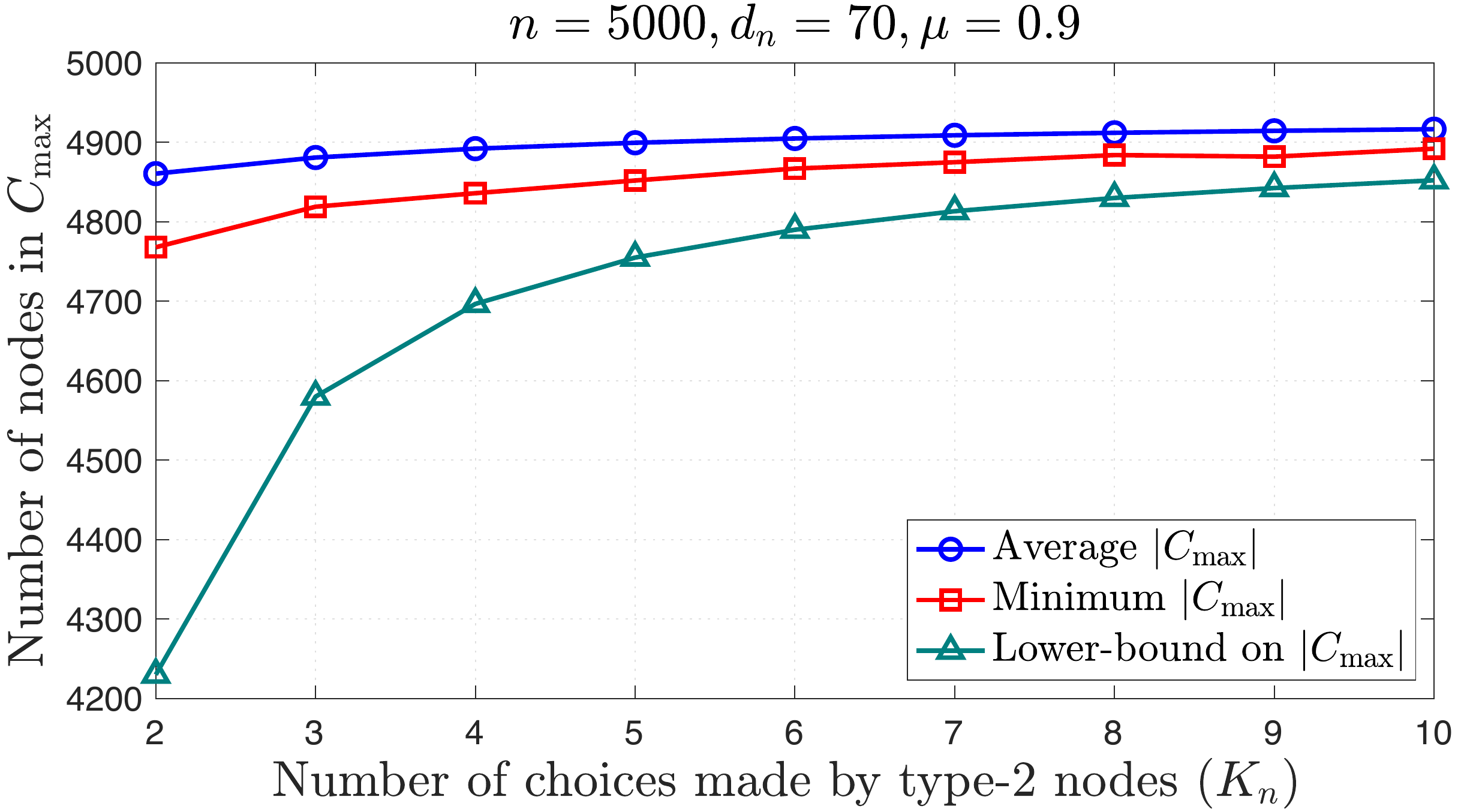}\label{fig:nodedel5000}
	\caption{\sl Average and minimum number of nodes contained in the largest connected component $(C_{\rm max})$ of $\hd$ across 100,000 independent experiments with (i) $n=1000$, $d_n=20$, $K_n=2$, and $\mu \in \{0.1, \dots, 0.9\}$ and (ii) $n=5000, d_n=70, \mu=0.9$, and $K_n \in \{2, \dots, 10\}$. We also plot the heuristic lower bound on the size of the largest connected component, as inferred from Theorem~\ref{theorem:gc-deleted}.}\label{fig:nodedel1000}
%		%We also plot a lower bound on the size of the giant component $(n - d_n/(\kk -1))$ inferred from Theorem~\ref{theorem:gc-deleted}.. 
%	} 
%%	\label{fig:gc_nodedel}
%\end{figure}
%\begin{figure}[!t]
%	\centering
%	\includegraphics[scale=0.34]{varymun1000dn20.pdf}\label{fig:nodedel1000}\vspace{5mm}
%	\includegraphics[scale=0.34]{varyK2n5000dn70.pdf}
%	\caption{\sl Average and minimum number of nodes contained in the largest connected component of $\hd$ across 100,000 independent experiments with $n=5000, \mu=0.9, d_n=70$, and $K_n \in \{2, \dots, 10\}$. The lower bound on the size of the giant component $(n - d_n/(\kk -1))$, as inferred from Theorem~\ref{theorem:gc-deleted}, stays below the experimentally observed minimum number of nodes across the 100,000 experiments. } \label{fig:nodedel5000}
%	\label{fig:gc_nodedel2}
\end{figure}

Next, we probe the size of the largest connected component of $\hd$, i.e., the graph obtained after the deletion of $d_n$ nodes uniformly at random from $\hh$. We first examine how the size of largest  connencted component of $\hd$ varies as we change $\mu$ or $\K$. We generate 100,000 independent realizations of $\hd$ with (i) $n=1000$, $d_n=20$, $K_n=2$, and vary $\mu \in \{0.1, \dots, 0.9\}$; and (ii) $n=5000, d_n=70, \mu=0.9$, and vary $K_n \in \{2, \dots, 10\}$.  For these parameters, we plot the {\em minimum} and average size of the largest component observed across the 100,000 realizations in Figure~\ref{fig:nodedel1000}. By comparing Figures~\ref{fig:gc_varymu} and~\ref{fig:nodedel1000} for $\K=2$ and $\mu=0.9$, we observe that for $n=1000$ (respectively, $n=5000$), after deleting $d_n=20$ (respectively, $d_n=70$) nodes, the maximum number of nodes observed outside $\cmax$ increases from 60 to 93 (respectively, 45 to 162). In Theorem~\ref{theorem:gc-deleted}, we characterized the asymptotic size of $\cmaxdn$, namely, when $|\cmaxdn| \geq n-d_n-x_n$ whp for any $x_n=\omega(1)$ satisfying $x_n> \frac{(1+\epsilon)d_n}{\kk-1}$  $\forall \epsilon>0$. Motivated by this, we propose a \emph{heuristic} lower bound on the size of the giant component given by $|\cmaxdn| \geq n - d_n- \frac{d_n}{\kk -1}$ for finite $n$ and we plot this in Figure~\ref{fig:nodedel1000}. We observe that this lower bound becomes tighter as $\mu$ is decreased or $\K$ is increased. Note that the minimum size of the largest connected component observed across 100,000 experiments stays above the corresponding lower bound for all values for $\mu$ and $\K$, reinforcing the usefulness of our results for finite node regimes.

%[Lower bound]%with $n=1000, d_n=20, \K=2$ and vary $\mu$ between 0.1 and 0.9 in increments of 0.1.  
%From Figure~\ref{fig:nodedel1000}, we can see that the minimum 

\section{Proof of Theorem~\ref{theorem:gc} and Corollary~\ref{cor:gc_2_classes}}
\subsection{Proof of Theorem~\ref{theorem:gc}}
\label{sec:proof}
In this section,  we provide a proof sketch for Theorem~\ref{theorem:gc}. %Recall that $\cmax$ denotes the largest connected component of $\hh$. %All details can be found in \cite{SoodYaganISIT2020_giant_comp}. 
%which implies that if $\K \geq 2$ for all $n$, then whp $\cmax$ contains all but finitely many of the nodes. 
%The proof of this result goes through a sequence of intermediate steps. 
We start  by  defining a {\em cut}.% as a subset of nodes that is isolated from the rest of the graph. 

\begin{definition}[Cut]
	\cite[Definition 6.3]{MeiPanconesiRadhakrishnan2008}
	Consider a graph $\mathcal{G}$ with the node set $\nodes$. A \emph{cut} is defined as a non-empty subset $S \subset \nodes$ of nodes 
	%of size at most $n/2$ 
	that is {\em isolated} from the rest of the graph. Namely,  $S \subset \nodes$ is a cut if there is no   edge between $S$ and $S\comp=\nodes \setminus S$.  
	\label{def:cut}
\end{definition}{}
It is clear from Definition~\ref{def:cut} that if $S$ is a cut, then so is $S\comp$.
It is important to note the distinction between a \emph{cut} as defined above and the notion of a \emph{connected component} given in Definition~\ref{def:concomp}. %A connected component is also a cut. However, 
%nodes within a cut may not be connected and every cut is a connected component.
A connected component is isolated from the rest of the nodes by Definition~\ref{def:concomp} and therefore it is also a cut. However, nodes within a cut may not be connected meaning that not every cut is a connected component. 
%{\color{red}TO DO: Add figures for proof outline (cut-sets) with/without node deletions}
Let $\mathcal{E}_n (\mu,K_n; S)$ denote the event that  $S \subset \nodes$ is a cut in $\hh$ as per Definition~\ref{def:cut}. The event $\mathcal{E}_n (\mu,K_n; S)$ occurs if no nodes in $S$ pick neighbors in $S\comp$, and no nodes in $S$ pick neighbors in $S\comp$. Thus, we have
\begin{align}
	\mathcal{E}_n (\mu,K_n; S) =
	\bigcap_{i \in S} \bigcap_{j \in S^c}
	\left(
	\left \{ i \not \in \Gamma_{n,j} \right \}
	\cap 
	\left \{ j \notin \Gamma_{n,i} \right \}
	\right). \nonumber
\end{align}

Let $\mathcal{Z}_n(x_n;\mu,K_n)$ denote the event that $\hh$ has no cut $S \subset \nodes$ with size  $x_n \leq |S| \leq n-x_n$ where  $x:\N_0 \rightarrow  \N_0$ is a sequence such that $x_n \leq {n}/{2} \ \forall n$. 
In other words, $\mathcal{Z}_n(x_n;\mu,K_n)$ is the event that there are no cuts in $\hh$ whose size falls in the range $[x_n, n-x_n]$. If $S$ is a cut, then so is $S^{c}$ (i.e., if there is a cut of size $m$ then there must be a cut of size $n-m$), therefore we see that  
\begin{align}
	\mathcal{Z}_n(x_n;\mu,K_n) & = \bigcap_{S \in \mathcal{P}_n: ~x_n\leq  |S| \leq \lfloor \frac{n}{2} \rfloor}  \left(\mathcal{E}_n({\mu},{K}_n; S)\right)\comp,
	\label{eq:gcintersection-nocut}
\end{align}
where $\mathcal{P}_n$ is the collection of all non-empty  subsets of $\nodes$.  
Taking the complement of both sides in (\ref{eq:gcintersection-nocut}) and then using a union bound we get
\begin{align}
	\pr\left[\left(\mathcal{Z}_n(x_n;\mu,K_n)\right)\comp\right] &\leq  \sum_{ S \in \mathcal{P}_n: x_n \leq |S| \leq \lfloor \frac{n}{2} \rfloor } \pr[ \mathcal{E}_n ({\mu},{K}_n; S) ] \nonumber \\
	&=\sum_{r=x_n}^{ \left\lfloor \frac{n}{2} \right\rfloor }
	\left ( \sum_{S \in \mathcal{P}_{n,r} } \pr[\mathcal{E}_n ({\mu},{K}_n; S)] \right ) \label{eq:BasicIdea+UnionBound},
\end{align}
where  $\mathcal{P}_{n,r} $ denotes the collection of all subsets of $\nodes$ with exactly $r$ elements.
For each $r=1, \ldots , n$, we simplify the notation by writing $\mathcal{E}_{n,r} ({\mu},{K}_n)=\mathcal{E}_n ({\mu},{K}_n ; \{ v_1, \ldots , v_r \} )$. From the exchangeability of the node labels and associated random variables, we get
\[
\pr[ \mathcal{E}_n({\mu},{K}_n ; S) ] = \pr[ \mathcal{E}_{n,r}(\mu,\K) ], \quad S \in
\mathcal{P}_{n,r}.
\]
Noting that $|\mathcal{P}_{n,r} | = {\binom{n}{r}}$, we obtain
\begin{equation}
	\sum_{S \in \mathcal{P}_{n,r} } \pr[\mathcal{E}_n ({\mu},{K}_n ; S) ] 
	= {\binom{n}{r}} ~ \pr[\mathcal{E}_{n,r} ({\mu},{K}_n)].  \nonumber
	\label{eq:ForEach=r}
\end{equation}
Substituting into (\ref{eq:BasicIdea+UnionBound}) we obtain 
\begin{align}
	\pr\left[\left(\mathcal{Z}_n(x_n;\mu,K_n)\right)\comp\right] \leq \sum_{r=x_n}^{ \left\lfloor \frac{n}{2} \right\rfloor }
	{\binom{n}{r}} ~ \pr[ \mathcal{E}_{n,r}({\mu},{K}_n) ] .
	\label{eq:gc_zbound}
\end{align}
Our next result presents an upper bound on $\pr\left[\left(\mathcal{Z}_n(M;\mu,K_n)\right)\comp\right]$, % It also shows that if $\K \geq 2$ for all $n$ and $0 < \mu < 1$, then a.a.s. $(\mathcal{Z}_n(x_n;\mu,K_n))\comp$ does {\em not} take place for any $x_n =\omega(1)$. In other words, %%$\mathcal{Z}_n(x_n;\mu,K_n))$ takes place a.a.s., meaning that 
i.e, the probability that there exists a cut with size in the range $[M,  n-M]$ for $\hh$. 
\begin{proposition}
{\sl	\label{prop:gcproofk3}
	Consider a scaling\footnote{We refer to any mapping $K:\N \rightarrow \N$ as a {\em scaling} if it satisfies the condition 
		\[
		2 \leq K_n < n, \quad  n=2, 3, \ldots .
		\] } $K: \N_0 \rightarrow \N_0$ such that $\K \geq 2 $  $\forall n$, $K_n = O(1)$, 
	%is bounded above for all $n$ 
	and $\mu \in (0,1)$. It holds that %Then for any sequence $x:\N_0 \rightarrow \N_0$ such that $x_n=\omega(1)$, we have
	\begin{align}
		& \pr\left[\left(\mathcal{Z}_n(M;\mu,K_n)\right)\comp\right]\nonumber \\
		&\leq \frac{\exp\{-M\left(\kk-1\right)(1-\oo(1))\}}{1-{\exp\{-\left(\kk-1\right)(1-\oo(1))\}}} + \oo(1) .\label{eq:gcpropsum}
	\end{align}{}}
\end{proposition}{}
The proof of Proposition~\ref{prop:gcproofk3} is presented in Section~\ref{sec:proof_prop}.
%Due to space limit, proof of Proposition~\ref{prop:gcproofk3} is given in \cite{SoodYaganISIT2020_giant_comp}.

The succeeding Lemma establishes the relevance of the event $ \mathcal{Z}_n(x_n;\mu,K_n)$ in obtaining a lower bound for the size of the largest connected component.  
\begin{lemma}
	{\sl
	\label{lem:gc_sum}
	For any sequence $x:\N_0 \rightarrow \N_0$ such that $x_n \leq \lfloor n/3 \rfloor $ for all $n$, we have
	\begin{align*}
		\mathcal{Z}_n(x_n;\mu,K_n) \implies    |\cmax| > n -x_n  . %\ \forall n
	\end{align*}{}}
\end{lemma}{}
%Lemma \ref{lem:gc_sum} shows that the occurrence of event $\mathcal{Z}_n(x_n;\mu,K_n))$ in turn implies that $|\cmax| \geq n -x_n$. 
%\vspace{-3mm}
\emph{Proof of Lemma~\ref{lem:gc_sum}}\\
Assume that 
$\mathcal{Z}_n(x_n;\mu,K_n)$
takes place, i.e.,  
there is no cut in $\hh$ of size in the range $\left[x_n,n-x_n\right]$. Since a connected component is also a cut, this also means that 
%first argue that the non-existence of a \emph{cut} of size in the range $[x_n,{\floor{n/2}}]$ whp implies that
there is no connected component of size in the range $[x_n,n-x_n]$. Since every graph has at least one connected component, it either holds that the largest one has size $|\cmax| > n-x_n$, or that 
$|\cmax| < x_n$. 
%Suppose for a contradiction that there is a connected component $\dub \subset \nodes$ such that $|\dub|\in[x_n,n-x_n]$ whp. Now, if $ |\dub|\in[x_n,\floor{n/2}]$, we note that $\dub$ itself is a cut of size in the range $[x_n,{\floor{n/2}}]$ whp. Otherwise if $|\dub|\in[\floor{n/2},n-x_n]$, then the set $\nodes \setminus \dub$ is a cut of size in the range $[x_n,{\floor{n/2}}]$. This contradicts the fact that  there is no cut of size in the range $[x_n,{\floor{n/2}}]$. Thus, there is no {connected component} of size in the range $[x_n,n-x_n]$. 
We now show that it must be the case that  $|\cmax|> n-x_n$ under the assumption that $x_n \leq  n/3$. %{\color{red} why this is WLOG in analysis}
%We have already proved that there is no {connected component} of size in the range $[x_n,n-x_n]$. All that we have to show to establish the lemma is that if $|\cmax|<x_n$, then we get a contradiction.
Assume towards a contradiction that $|\cmax|<x_n$
meaning that 
the size of each connected component is less than $x_n$. Note that the union of any set of connected components is either a cut, or it spans the entire network. If no cut exists with size in the range $[x_n, n-x_n]$, then the union of any set of connected components should also have a size outside of $[x_n, n-x_n]$. Also, the union of all connected components has size $n$.  Let $C_1, C_2, \ldots, C_{\text{max}}$ denote the set of connected components in increasing size order. Let $m \geq 1$ be the largest integer such that $\sum_{i=1}^m |C_i| < x_n$. Since $|C_{m+1}| < x_n$, we have
\[
x_n \leq \sum_{i=1}^{m+1} |C_i| < x_n+x_n \leq 2n/3 \leq n -x_n.
\]
This means that $\cup_{i=1}^{m+1} C_i$ constitutes a cut with size
in the range $[x_n, n-x_n]$
contradicting the event $\mathcal{Z}_n(x_n;\mu,K_n)$. 
We thus conclude that if $\mathcal{Z}_n(x_n;\mu,K_n)$ takes place with $x_n \leq n/3$, then we must have
$|\cmax|> n-x_n$.
\myendpf

We now have all the requisite ingredients for establishing Theorem~\ref{theorem:gc}. Substituting $x_n=M, ~ \forall n$ in Lemma~\ref{lem:gc_sum} 
for any positive integer $M$ satisfying $M \leq n/3$,  
we get 
$$     \mathcal{Z}_n(M;\mu,K_n) \implies    |\cmax| > n -M.$$
%\nonumber, %\ \forall n 
%\end{align}{}
Equivalently, we have
$
|\cmax| \leq n -M  \implies \mathcal{Z}_n(M;\mu,K_n) \comp  %\label{eq:gcl1}
$. This gives for all $M \leq n/3$, 
\begin{equation}
	\pr\left[{|\cmax| \leq n -M} \right ] \leq \pr\left[{\mathcal{Z}_n(M;\mu,K_n) \comp}\right]
	\label{eq:gcl1_new}
\end{equation}
Combining (\ref{eq:gcl1_new}) with Proposition~\ref{prop:gcproofk3},  we get for all $M\leq n/3$, 		\begin{align}
	&  \pr \left[ |\cmax| \leq n-  M\right] \nonumber \\
	&\leq \frac{\exp\{-M\left(\kk-1\right)(1-\oo(1))\}}{1-{\exp\{-\left(\kk-1\right)(1-\oo(1))\}}} + \oo(1) 
\end{align}
Note that for all $M, N \in \N_0$ such that $M\leq N$, we have $|\cmax|  \leq  n-N \implies  |\cmax|  \leq  n-M$. This implies for each $M = 1, 2, \ldots$, the upper bound (\ref{eq:gc_theorem}) holds true, completing the proof of Theorem~\ref{theorem:gc}. %{\color{red}*to check*}
%
%%sequences $~ y:\N_0\rightarrow\N_0$ such that $x_n \leq y_n ~\forall n$, we have $n- |\cmax|  \leq  x_n  \implies  n- |\cmax|  \leq  y_n$ and therefore 
%
%
%Noting that 
%%and we get (\ref{eq:gc_theorem}) for all $M\leq n/3$. Note that
%%Proposition~\ref{prop:gcproofk3} in (\ref{eq:gcl1_new}). 
\myendpf
\subsection{Proof of Corollary~\ref{cor:gc_2_classes}}
Consider an arbitrary sequence $x_n=\omega(1)$. Substituting $M$ with $x_n$ in (\ref{eq:gc_theorem}), we readily see that 
%the number of nodes outside the largest connected component,  given by $n - |\cmax|$, satisfies
% \begin{align}
%   \limit  \pr \left[ |\cmax| < n-  x_n\right]=0
% \end{align}{}
\begin{align}
	\limit  \pr \left[  n- |\cmax|  \leq  x_n\right]=1. \label{eq:cor-initial-proof}
\end{align}{}
Namely, we have 
\begin{align}
	n - |\cmax| \leq x_n \ {\rm whp~for~any} \ \ x_n =\omega(1).
	\label{eq:gc_xn_whp}
\end{align}
This is equivalent to the number of nodes ($n - |\cmax|$) outside the largest connected component being  {\em bounded}, i.e., $O(1)$, with high probability. This fact is sometimes stated using the probabilistic big-O notation, $O_p$. A random sequence $f_n = O_p(1)$ if for any $\varepsilon>0$ there exists finite integers $M(\varepsilon)$ and $n(\varepsilon)$ such that
$\pr[f_n > M(\varepsilon)] < \varepsilon$ for all $n \geq n(\varepsilon)$.
In fact, we see from {\cite[Lemma~3]{janson2011probability}%\footnote{how do additional restrictions on $x_n$ impact this?}
} that (\ref{eq:gc_xn_whp}) is equivalent to having
$  n - |\cmax| = O_p(1)$
%. 
%\label{eq:gc_xn_Op}  
%\end{align}
Here, we equivalently state this as 
\begin{align}{}
	n - |\cmax| = O(1) \ \ {\rm whp}, \nonumber
\end{align}
giving readily (\ref{eq:gc_xn_BigO}).
%{\color{blue}*to check*}
\label{subsec:cor1}
\myendpf

%\section*{Appendix}

\section{Proof of Proposition~\ref{prop:gcproofk3}}
\label{sec:proof_prop}
\subsection{Useful facts}
\noindent For $0 \leq x < 1$ and for a sequence $y = 0,1,2\dots,$ it is known\cite[Fact 2]{ZhaoYaganGligor} that
\begin{align}
	1-xy \leq  (1-x)^y \leq 1-xy+\frac{1}{2}x^2y^2.
	\label{eq:gc_junfact}
\end{align}
%{\color{blue}A proof of (\ref{eq:gc_junfact})} can be found
%in \cite[Fact 2]{ZhaoYaganGligor}. 
For all $x \in \mathbb{R}$, we have
\begin{align}
	1  \pm x  &\leq e^{\pm x}. \label{eq:gc_1pmx}
\end{align}
For $0\leq m \leq n_1 \leq n_2, ~~  m,n_1,n_2 \in \N_0$,
\begin{align}
	\frac{{\binom{n_1}{m}}}{{\binom{n_2}{m}}}=\prod\limits_{i=0}^{m-1} \left( \frac{n_1-i}{n_2-i}\right) \leq \left( \frac{n_1}{n_2}\right)^m. \label{eq:gc_choosefrac}
\end{align}
From \cite[Fact 4.1]{eletrebycdc2018}, we have that for $r=1,2,\dots, \lfloor \frac{n}{2} \rfloor$,
\begin{align}
	{\binom{n}{r}}& \leq \left(\frac{n}{r} \right)^r  \left(\frac{n}{n-r} \right)^{n-r}.
	\label{eq:kcon_cdc18fact}\end{align}
For $d_n \leq n$, this gives
\begin{align}
	{\binom{n-d_n}{r}}& \leq \left(\dfrac{n-d_n}{r} \right)^r \left(\dfrac{n-d_n}{n-d_n-r} \right)^{n-d_n-r} \label{eq:cdc18fact-dn}.
\end{align}
\subsection{Proof of Proposition~\ref{prop:gcproofk3}}
%\myproof
%In view of (\ref{eq:gc_zbound}), 
%\myproof
%\begin{comment}

In view of (\ref{eq:gc_zbound}), the proof for Proposition~\ref{prop:gcproofk3} will follow upon showing
\begin{align}
	& \sum_{r=M}^{\floor{n/2}} {\binom{n}{r}}\pr[\sdcon]\nonumber \\
	&\leq \frac{\exp\{-M\left(\kk-1\right)(1-\oo(1))\}}{1-{\exp\{-\left(\kk-1\right)(1-\oo(1))\}}} + \oo(1) . \label{eq:propeqn}
\end{align}{}

We have
\begin{align}
	& {\binom{n}{r}}\pr[\sdcon]\nonumber \\
	&={\binom{n}{r}} \left( \mu \left(\dfrac{n-r-1}{n-1}\right)+(1-\mu) \dfrac{{\binom{n-r-1}{\K}}}{{\binom{n-1}{\K}}} \right)^{n-r}  \nonumber \\
	& \quad \cdot \left( \mu \left(\dfrac{r-1}{n-1}\right)+(1-\mu) \dfrac{{\binom{r-1}{\K}}}{{\binom{n-1}{\K}}} \right)^{r} \nonumber  \\
	&\leq{\binom{n}{r}}\left( \mu \left(1-\dfrac{r}{n-1}\right)+
	(1-\mu) \left(1-\dfrac{r}{n-1}\right)^{\K} \right)^{n-r}\nonumber \\
	&  \quad \cdot \left( \mu \left(\dfrac{r-1}{n-1}\right)+(1-\mu) \left(\dfrac{{r-1}}{{n-1 }} \right)^{\K} \right)^{r} \label{eq:gc_eq1} \\
	&\leq{\binom{n}{r}}\left( \mu \left(1-\dfrac{r}{n}\right)+
	(1-\mu) \left(1-\dfrac{r}{n}\right)^{\K} \right)^{n-r}\nonumber \\
	&  \quad \cdot \left( \mu \left(\dfrac{r}{n}\right)+(1-\mu) \left(\dfrac{{r}}{{n }} \right)^{\K} \right)^{r}  \nonumber  \\
	& \leq  \left(\frac{n}{r} \right)^r  \left(\frac{n}{n-r} \right)^{n-r}\left(1-\dfrac{r}{n}\right)^{n-r} \left( \dfrac{r}{n}\right)^r \nonumber \\
	& \quad \cdot \left( \mu+
	(1-\mu) \left(1-\dfrac{r}{n}\right)^{\K-1} \right)^{n-r}\nonumber \\
	& \quad \cdot \left( \mu +(1-\mu) \left(\dfrac{{r}}{{n }} \right)^{\K-1} \right)^{r} \label{eq:gc_eq2} \\
	& = \left( \mu+
	(1-\mu) \left(1-\dfrac{r}{n}\right)^{\K-1} \right)^{n}\nonumber \\
	& \quad \cdot \left(  \dfrac{\mu +(1-\mu) \left(\dfrac{{r}}{{n }} \right)^{\K-1}}{ \left( \mu+
		(1-\mu) \left(1-\dfrac{r}{n}\right)^{\K-1} \right)} \right)^{r} \nonumber \\
	& \leq  \left( \mu+
	(1-\mu) \left(1-\dfrac{r}{n}\right)^{\K-1} \right)^{n} \label{eq:gc_beforesplit}
\end{align}
where (\ref{eq:gc_eq1}) uses (\ref{eq:gc_choosefrac}), (\ref{eq:gc_eq2}) follows from (\ref{eq:kcon_cdc18fact}) and (\ref{eq:gc_beforesplit}) is plain from the observation that $ r/n \leq 1/2$.
\par We divide the summation in (\ref{eq:propeqn}) into two parts depending on whether $r$ exceeds ${n}/{\log n}$. The steps outlined below can be used to upper bound the summation in (\ref{eq:propeqn}) for an arbitrary splitting of the summation indices.\\
\begin{align}
	\sum_{r=M}^{\floor{n/2}} {\binom{n}{r}}\pr[\sdcon]\nonumber&=  \sum_{r=M}^{\floor{n/\log n}} {\binom{n}{r}}\pr[\sdcon]\nonumber \\
	&+\sum_{r=\floor{n/\log n}}^{\floor{n/2}} {\binom{n}{r}}\pr[\sdcon].\label{eq:gcsplit} \end{align}{}
We first upper bound each term in the summation with indices in the range $M  \leq r \leq \floor{n/\log n}$.\\

\noindent \textbf{ Range 1:  $M  \leq r \leq \floor{n/\log n}$}\\ 
\begin{align}
	&{\binom{n}{r}}\pr[\sdcon]  \nonumber \\
	&\leq \left( \mu+(1-\mu) \left(1-\dfrac{r}{n}\right)^{\K-1} \right)^{n}  \nonumber\\
	& =  \left( 1- (1-\mu) \left(1-\left(1-\dfrac{r}{n}\right)^{\K-1}\right) \right)^{n} \label{eq:gc-del-case1-init}
	%% \leq  \left(1-\dfrac{r}{n}\right)^{1} \right)
\end{align}
For $r  \leq \floor{n/\log n}$, we have {$\frac{r}{n}=\oo(1)$}. Using Fact (\ref{eq:gc_junfact}) with $x=\frac{r}{n}$ we get
\begin{align}
	&{\binom{n}{r}}\pr[\sdcon]  \nonumber \\
	&\leq \left( 1- (1-\mu)\left(1-\left(1-\dfrac{r(\K-1)}{n}+\dfrac{r^2 (\K-1)^2}{2n^2} \right)\right) \right)^{n} \nonumber \\
	%&= \left( 1- (1-\mu)\left(\dfrac{r(\K-1)}{n}-\dfrac{r^2 (\K-1)^2}{2n^2} \right) \right)^{n}  \nonumber \\
	&= \left( 1- (1-\mu)\dfrac{r(\K-1)}{n}\left(1-\dfrac{r (\K-1)}{2n} \right) \right)^{n}  \label{eq:gc_range1-1}
\end{align}
Using $r \leq n / \log n$, (\ref{eq:gc_1pmx}) and that $\K=\OO(1)$, we obtain,

\begin{align}
	&{\binom{n}{r}}\pr[\sdcon]  \nonumber \\
	&\leq \left( 1- (1-\mu)\dfrac{r(\K-1)}{n}\left(1-\dfrac{ (\K-1)}{2 \log n} \right) \right)^{n}   \nonumber \\
	& \leq \exp \left\{-(1-\mu){r(\K-1)}\left(1-\dfrac{ (\K-1)}{2 \log n} \right)\right\} \nonumber \\ 
	& = \exp \left\{-r(1-\mu){(\K-1)}\left(1- \oo(1) \right)\right\}  \nonumber\\ 
	& = \exp \left\{-{r(\kk-1)}\left(1- \oo(1) \right)\right\}.   \label{eq:gc_range1}
\end{align}
Next, we upper bound the second term in the summation (\ref{eq:gcsplit}) with indices in the range $\floor{n/\log n}+1 \leq r \leq {\floor{n/2}}$.
\\\noindent \textbf{ Range 2: $\floor{n/\log n}+1 \leq r \leq {\floor{n/2}}$}\\
%Using $\K \geq 2$ and ($\ref{eq:gc_1pmx}$) we have
Observe that
\begin{align}
	{\binom{n}{r}}\pr[\sdcon]& \leq
	\left( \mu+(1-\mu) \left(1-\dfrac{r}{n}\right)^{\K-1} \right)^{n}\nonumber \\
	& \leq
	\left( \mu+(1-\mu) \left(1-\dfrac{r}{n}\right) \right)^{n} \label{eq:gc_range2b} \\
	& =
	\left( 1- \dfrac{r}{n} (1-\mu)\right)^{n} \nonumber \\
	& \leq \exp \left\{-r(1-\mu)\right\} , \label{eq:gc_range2a}
	%\\
	% & =\oo(1), \label{eq:gc_range2}
	%% \leq  \left(1-\dfrac{r}{n}\right)^{1} \right)
\end{align}
where (\ref{eq:gc_range2a}) follows from noting that $\K \geq 2$ and (\ref{eq:gc_range2b}) is a consequence of (\ref{eq:gc_1pmx}). Finally, we use (\ref{eq:gc_range1}) and (\ref{eq:gc_range2a}) in (\ref{eq:gcsplit}) as follows.
\begin{align}
	& \sum_{r=M}^{\floor{n/2}} {\binom{n}{r}}\pr[\sdcon]\nonumber \\
	&=  \sum_{r=M}^{\floor{n/\log n}} {\binom{n}{r}}\pr[\sdcon] \nonumber \\
	&+ \sum_{r=\floor{n/\log n}+1}^{\floor{n/2}} {\binom{n}{r}}\pr[\sdcon] \nonumber\\
	&\leq  \sum_{r=M}^{\floor{n/\log n}} \hspace{-2mm}  e^{-{r(\kk-1)}\left(1- \oo(1) \right)} + \hspace{-2mm} \sum_{r=\floor{n/\log n}+1}^{\floor{n/2}} \hspace{-2mm} e^{-r(1-\mu)}\nonumber\\
	&\leq  \sum_{r=M}^{\infty}  e^{-{r(\kk-1)}\left(1- \oo(1) \right)} + \hspace{-2mm} \sum_{r=\floor{n/\log n}+1}^{\infty} \hspace{-3mm} e^{-r(1-\mu)}\nonumber.
	% & \leq \left(  \sum_{r=M}^{\floor{n/\log n}}  \exp \left\{-{r(\kk-1)}\left(1- \oo(1) \right)\right\}\right) + e^{-(1-\mu)\frac{n}{\log n}}  \nonumber \\
	%  &\leq \left(  \sum_{r=M}^{\infty}  \exp \left\{-{r(\kk-1)}\left(1- \oo(1) \right)\right\}\right) + \oo(1).  \nonumber
\end{align}
Observe that both of the above geometric series have all terms strictly less than one, and thus they are {\em summable}. This gives
{
	\begin{align}
		&\sum_{r=M}^{\floor{n/2}}{\binom{n}{r}}\pr[\sdcon] \nonumber \\
		&\leq \frac{e^{-M\left(\kk-1\right)(1-\oo(1))}}{1-{e^{-\left(\kk-1\right)(1-\oo(1))}}} + \frac{e^{-(1-\mu)\frac{n}{\log n}}}{1-e^{-(1-\mu)}}
		\nonumber    \\
		&= \frac{e^{-M\left(\kk-1\right)(1-\oo(1))}}{1-{e^{-\left(\kk-1\right)(1-\oo(1))}}} + \oo(1).
		\nonumber
\end{align}}
\hfill\fsquare

% \begin{align}
% \E\left[\sum_{j\in \nodes_{-i}}\nonumber \ii\{i \sim j\}\right] &= (n-1)\pr[i \sim j],\\
% &=2\kk-\dfrac{\kk^2}{n-1}. \label{eq:avgnd}
% \end{align}

 \section{Proof of Theorem~\ref{theorem:gc-deleted}  and Corollary~\ref{cor:gc-deleted}}
 %{\color{red}*to check*}
 \label{sec:proof-deleted}
 \subsection{Proof of Theorem~\ref{theorem:gc-deleted}}
 \label{sec:proof-thm-del}
We extend the framework developed in Section~\ref{sec:proof} to analyze the giant component in $\hd$. Recall that in Section~\ref{sec:model}, we defined $\hd$ as the subgraph of $\hh$ induced by the subset of nodes contained in $\nodes \setminus D$, where $D \subset \nodes, |D|=d_n$ and $D$ is selected uniformly at random from all possible subsets of $\nodes$ with cardinality $d_n$. 
Let $\ssdcondn$ denote the event that $S \subset \nodes \setminus D$ is a cut in $\hd$ as per Definition~\ref{def:cut}. The event $\ssdcondn$ occurs if nodes in $S$ pick neighbors in $S \cup D$, and nodes in $S\comp$ pick neighbors in $S\comp \cup D$. Let $\zdn$ denote the event that $\hd$ contains no cut of size in the range $[x_n , n-d_n - x_n]$. From Lemma \ref{lem:gc_sum}, it follows that if $\zdn$ occurs, then the size of the largest connected component of $\hd$ is greater than $n - d_n - x_n$ whp. We have
\begin{align}
	&\zdn \nonumber\\& = \bigcap_{S \in \mathcal{P}_n: ~x_n\leq  |S| \leq \lfloor \frac{n-d_n}{2} \rfloor}  \left(\ssdcondn\right)\comp \nonumber,
\end{align}
where $\mathcal{P}_n$ is the collection of all non-empty subsets of $ \nodes \setminus D$. Using a union bound, we get 
\begin{align}
	&\pr[\zdn\comp] \nonumber\\
	&\leq \sum_{ S \in \mathcal{P}_n: x_n \leq |S| \leq \lfloor \frac{n-d_n}{2} \rfloor } \pr[ \ssdcondn ] \nonumber \\
	&=\sum_{r=x_n}^{ \left\lfloor \frac{n-d_n}{2} \right\rfloor }
	\left ( \sum_{S \in \mathcal{P}_{n,r} } \pr[\ssdcondn] \right ), \label{eq:gc_zcondcon}
\end{align}
where  $\mathcal{P}_{n,r} $ denotes the collection of all subsets of $\nodes\setminus D$ with exactly $r$ elements.
For each $r=1, \ldots , \left\lfloor (n-d_n)/2\right\rfloor$, let $\sdcondn=\mathcal{E}_n ({K}_n,\mu,{d_n} ; \{v_1, \ldots , v_r \} )$. From the exchangeability of the associated random variables, we have $\forall S \in \mathcal{P}_{n,r} $,
\begin{align}
\pr[ \ssdcondn ] = \pr[ \sdcondn ]. \nonumber
\end{align}
Substituting in (\ref{eq:gc_zcondcon}), we get 
\begin{align}
	&\pr[\zdn\comp] \nonumber\\
	&\leq \sum_{r=x_n}^{ \left\lfloor \frac{n-d_n}{2} \right\rfloor }
	\left ( \sum_{S \in \mathcal{P}_{n,r} } \pr[\sdcondn] \right ) \nonumber\\
	&=\sum_{r=x_n}^{\floor{(n-d_n)/2}} {\binom{n-d_n}{r}}{\pr[\sdcondn]} \label{eq:gc-del-sum}.
\end{align}

Next, we find conditions on $x_n$ under which the summation (\ref{eq:gc-del-sum}) decays to zero.  In the above summation (\ref{eq:gc-del-sum})  we seek to derive an upper bound for the event $\left(\mathcal{Z}_n(x_n;\mu,K_n)\right)\comp$, which through Lemma \ref{lem:gc_sum} gives an upper bound on the event that $x_n$ or more nodes lie outside the largest connected component of $\hd$. As specified in Theorem~\ref{theorem:gc-deleted}, throughout, we assume that $x_n > \frac{(1+\epsilon)d_n}{\kk-1} $ and $x_n=o(n)$.  Therefore, for a given $d_n$ such that $d_n=o(n)$, our goal is to find the smallest $x_n$ %(such that $x_n=\omega(1)$ that satisfies (\ref{eq:gc-del:infsuffcond}) and 
that drives the summation in (\ref{eq:gc-del-sum}) to $o(1)$.  However, note that there is a trade-off while chosing $x_n$. If we choose a smaller $x_n$ that drives the summation in (\ref{eq:gc-del-sum}) to $o(1)$, although we are guaranteed a larger giant component whp through Lemma \ref{lem:gc_sum}, we are including more terms in the summation and the resulting upper bound on the summation can decay at a slower rate.  The Proposition below gives an upper bound on $\pr\left[\left(\zdn \right)\comp\right]$. % It also shows that if $\K \geq 2$ for all $n$ and $0 < \mu < 1$, then a.a.s. $(\mathcal{Z}_n(x_n;\mu,K_n))\comp$ does {\em not} take place for any $x_n =\omega(1)$. In other words, %%$\mathcal{Z}_n(x_n;\mu,K_n))$ takes place a.a.s., meaning that 
%i.e, the probability that there exists a cut with size in the range $[M,  n-M]$ for $\hh$. 
\begin{proposition}
	{\sl \label{prop:gc-del}
%		{\sl For the graph $\hd$ with \\$d_n=o(n)$, $K_n \geq 2 ~\forall n$ and $K_n =O(1)$. For any sequence $x:\N_0 \rightarrow \N_0$,\\
%		if $x_n > \dfrac{d_n}{(\kk -1 )}~\forall n$
%		, then
	Consider a scaling $K: \N_0 \rightarrow \N_0$ such that $\K \geq 2 $  $\forall n$, $K_n = O(1)$, 
	%is bounded above for all $n$ 
	and $\mu \in (0,1)$ with $d_n=o(n)$. For any $x:\N_0 \rightarrow \N_0$ such that $x_n > \frac{(1+\epsilon)d_n}{\kk -1 }~\forall n$ and $x_n=o(n)$, we have%Then for any sequence $x:\N_0 \rightarrow \N_0$ such that $x_n=\omega(1)$, we have
	\begin{align}
		& \pr\left[\left(\zdn \right)\comp\right]\nonumber \\
	%	&\leq  \frac{e^{-\frac{x_n}{2}\left(\kk-1\right)(1-\oo(1))}}{1-{e^{-\left(\kk-1\right)(1-\oo(1))}}} + \frac{e^{-\frac{x_n}{2}\left(1-\mu\right)(1-\oo(1))}}{1-{e^{-\left(1-\mu\right)(1-\oo(1))}}}. \nonumber		
		&\leq \frac{e^{- x_n (\kk-1) \frac{\epsilon}{1+\epsilon}(1-\oo(1))}}{1-{e^{-\left(\kk-1\right)\frac{\epsilon}{1+\epsilon}(1-\oo(1))}}} + \frac{e^{-{x_n}\left(1-\mu\right)\frac{\epsilon}{1+\epsilon}(1-\oo(1))}}{1-{e^{-\left(1-\mu\right)\frac{\epsilon}{1+\epsilon}(1-\oo(1))}}}. \nonumber
	\end{align}{}}
\end{proposition}{}
The proof of Proposition~\ref{prop:gc-del} is presented in Section~\ref{sec:proof_prop_deleted}. We present an alternate upper bound under the stronger condition that $x_n >\frac{(1+\epsilon)d_n}{1-\mu }~\forall n$ in the Appendix.
 %{\color{blue}remark- trade-off $x_n$}

Invoking Lemma~\ref{lem:gc_sum} for the graph $\hd$, we have for any sequence $x:\N_0 \rightarrow \N_0$ such that $x_n \leq \lfloor (n-d_n)/3 \rfloor $ for all $n$, we have
	\begin{align}
	\zdn \implies    |\cmaxdn| > n -d_n -x_n,  \label{eq:gc-del-lem-implication}
	%\ \forall n 
\end{align}{}
or equivalently,
	\begin{align}
	   & \pr[|\cmaxdn| \leq n -d_n -x_n] \nonumber\\
	    &\quad \leq  \pr\left[\left(\zdn \right)\comp\right].\label{eq:gc-del-lem}  %\label{eq:gc-del-lem} 
	%\ \forall n 
\end{align}{}
Combining \eqref{eq:gc-del-lem} with Proposition~\ref{prop:gc-del}, we get for all $:\N_0 \rightarrow \N_0 $ such that $\frac{(1+\epsilon)d_n}{\kk -1 }<x_n \leq \lfloor (n-d_n)/3 \rfloor~\forall n$ and $x_n=o(n)$, 
	\begin{align}
	&  \pr[|\cmaxdn| \leq n -d_n -x_n] \nonumber\\%\pr\left[\left(\zdn \right)\comp\right]\nonumber \\
	%	&\leq  \frac{e^{-\frac{x_n}{2}\left(\kk-1\right)(1-\oo(1))}}{1-{e^{-\left(\kk-1\right)(1-\oo(1))}}} + \frac{e^{-\frac{x_n}{2}\left(1-\mu\right)(1-\oo(1))}}{1-{e^{-\left(1-\mu\right)(1-\oo(1))}}}. \nonumber		
	&\leq \frac{e^{- x_n (\kk-1) \frac{\epsilon}{1+\epsilon}(1-\oo(1))}}{1-{e^{-\left(\kk-1\right)\frac{\epsilon}{1+\epsilon}(1-\oo(1))}}} + \frac{e^{-{x_n}\left(1-\mu\right)\frac{\epsilon}{1+\epsilon}(1-\oo(1))}}{1-{e^{-\left(1-\mu\right)\frac{\epsilon}{1+\epsilon}(1-\oo(1))}}}. \nonumber
\end{align}{}
For arbitrary sequences $~ y:\N_0\rightarrow\N_0$ such that $x_n \leq y_n ~\forall n$, we have $ |\cmaxdn|  \leq  n-d_n-y_n  \implies  |\cmaxdn|  \leq  n-d_n-x_n$ and therefore for all sequences $x_n$ such that $x_n > \frac{(1+\epsilon)d_n}{\kk -1 }~\forall n$, the upper bound (\ref{eq:gc_general_theorem}) holds true; this completes the proof of Theorem~\ref{theorem:gc-deleted}. 

\hfill\fsquare
%{\color{blue}proof}
\subsection{Proof of Corollary~\ref{cor:gc-deleted}}
\noindent {i)} When $d_n =O(1)$, the condition $x_n > \frac{(1+\epsilon)d_n}{\kk -1 }$ is satisfied for sufficiently large $n$ for any $x_n=\omega(1)$. Substituting $x_n=\omega(1)$ in (\ref{eq:gc_general_theorem}), we get for $d_n=O(1)$, 
\begin{align}
	\limit  \pr \left[  n-d_n- |\cmaxdn|  \leq  x_n\right]=1.\nonumber
\end{align}{}
Thus, for $d_n =O(1)$, we have for any $x_n=\omega(1)$ that
\begin{align}
	n - d_n- |\cmaxdn| \leq x_n \ {\rm whp}, \nonumber
\end{align}
which is equivalent to the number of nodes outside the largest connected component in $\hd$ being {\ bounded}, i.e., $n-d_n-|\cmaxdn|=O(1)$ whp.  Since, $d_n$ itself is bounded, we get $|\cmaxdn|=n-O(1)$ whp.\\
\noindent {ii)} We have already analyzed the case  $d_n=O(1)$ above. Now, we consider the case $d_n =o(n)$ and $d_n=\omega(1)$. Using (\ref{eq:gc_general_theorem}), for any $x_n=\omega(1)$ satisfying $x_n> \frac{(1+\epsilon)d_n}{\kk -1 }$, we have
\begin{align}
	\limit  \pr \left[  n-d_n- |\cmaxdn|  \leq  x_n\right]=1,\nonumber
\end{align}{}
Upon selecting $x_n = \frac{(1+\epsilon)d_n}{\kk -1 } + \delta >0 ~\forall n$ where $\delta>0$, note that $x_n$ satisfies both $x_n=\omega(1)$ and $x_n> \frac{(1+\epsilon)d_n}{\kk -1 }$ for $d_n=o(n)$ and $d_n=\omega(1)$, yielding
%or equivalently, for any  $x_n=\omega(1)$ satisfying $x_n> \frac{(1+\epsilon)d_n}{\kk -1 }$,
\begin{align}
 |\cmaxdn|  \geq 	n - d_n- \frac{(1+\epsilon)d_n}{\kk -1 }-\delta\ {\rm whp}.\nonumber
\end{align}
Therefore, for $d_n=o(n)$, it holds that $|\cmaxdn| = n(1-o(1))$ whp.
\hfill\fsquare
\section{Proof of Proposition \ref{prop:gc-del}}
\label{sec:proof_prop_deleted}
\subsection{Preliminaries}
In order to find a vanishing upper bound on summation \eqref{eq:gc-del-sum}, we first upper bound the individual terms as follows.
\begin{align}
	& {\binom{n-d_n}{r}}\pr[\sdcondn]\nonumber\\ 
	&={\binom{n-d_n}{r}} 
	\left( \mu \left(\dfrac{r+d_n-1}{n-1}\right)+(1-\mu) \dfrac{{\binom{r+d_n-1}{\K}}}{{\binom{n-1}{\K}}} \right)^{r}
	\nonumber \\
	&\quad \cdot  \left( \mu \left(\dfrac{n-r-1}{n-1}\right)+(1-\mu) \dfrac{{\binom{n-r-1}{\K}}}{{\binom{n-1}{\K}}} \right)^{n-d_n-r} \nonumber  \\
	%%%
	&\leq{\binom{n-d_n}{r}} 
	\left( \mu \left(\dfrac{r+d_n-1}{n-1}\right)+(1-\mu)  \left(\dfrac{r+d_n-1}{n-1}\right)^{K_n} \right)^{r}
	\nonumber\\
	&\quad \cdot  \left( \mu \left(\dfrac{n-r-1}{n-1}\right)+(1-\mu) \left(\dfrac{n-r-1}{n-1}\right)^{\K} \right)^{n-d_n-r} \label{eq:gc-del-1}  \\
	%%%%%
	&\leq {\binom{n-d_n}{r}} \left(\dfrac{r+d_n}{n}\right)^r\left( \mu +(1-\mu) \left(\dfrac{r+d_n}{n}\right)^{\K-1} \right)^{r}\nonumber
	\\
	&\quad \cdot \left(\dfrac{n-r}{n}\right)^{n-d_n-r} \left( \mu +(1-\mu) \left(\dfrac{n-r}{n}\right)^{\K-1} \right)^{n-d_n-r} 	\label{eq:gc-del-2} \\%%
	%%%%%
	&={\binom{n-d_n}{r}} \left(\dfrac{r+d_n}{n}\right)^r  \left(\dfrac{n-r}{n}\right)^{n-d_n-r} 
	\nonumber\\
	&\quad \cdot \left( \mu +(1-\mu) \left(\dfrac{n-r}{n}\right)^{\K-1} \right)^{n-d_n} \nonumber  \\%%
	&\quad \cdot \left(  \dfrac{\mu +(1-\mu) \left(\dfrac{{r+d_n}}{{n }} \right)^{\K-1}}{  \mu+
		(1-\mu) \left(\dfrac{n-r}{n}\right)^{\K-1} } \right)^{r}  	\nonumber \\
	%%%%%
	&\leq {\binom{n-d_n}{r}} \left(\dfrac{r+d_n}{n}\right)^r  \left(\dfrac{n-r}{n}\right)^{n-d_n-r} \nonumber
	\\
	&\quad \cdot \left( \mu +(1-\mu) \left(1-\dfrac{r}{n}\right)^{\K-1} \right)^{n-d_n},   	\label{eq:gc-del-3}
\end{align}
where \eqref{eq:gc-del-1} follows from \eqref{eq:gc_choosefrac},  \eqref{eq:gc-del-2} follows from noting that $r \leq  (n-d_n)/2 \implies r+ d_n \leq n-r <n$
and \eqref{eq:gc-del-3} follows from
%from the fact that $r \leq  (n-d_n)/2$ and thus $\frac{{r+d_n}}{{n }} \leq  1-\frac{r}{n}$, combined with 
the observation that $\mu +(1-\mu)x^{\K -1}$ is an increasing function in $x$ where $x>0$ combined with the fact that $r+ d_n \leq n-r$. 
%\subsection{Some useful bounds}
We now prove a bound to simplify \eqref{eq:gc-del-3}.
\begin{align}
	&{\binom{n-d_n}{r}} \left(\dfrac{r+d_n}{n}\right)^r  \left(\dfrac{n-r}{n}\right)^{n-d_n-r} \nonumber\\ 
	& \leq \left(\dfrac{n-d_n}{r} \right)^r \left(\dfrac{n-d_n}{n-d_n-r} \right)^{n-d_n-r} \nonumber\\
	&\qquad \cdot \left(\dfrac{r+d_n}{n}\right)^r \left(\dfrac{n-r}{n}\right)^{n-d_n-r}\label{eq:gc-del-4} \\
	&=\left(\dfrac{n-d_n}{n} \right)^r \left(\dfrac{r+d_n}{r}\right)^r \nonumber\\
	&\qquad \cdot \left(\dfrac{n-d_n}{n}\right)^{n-d_n-r}  \left(\dfrac{n-r}{n-d_n-r} \right)^{n-d_n-r}	\nonumber\\
	&=\left(\dfrac{r+d_n}{r}\right)^r  \left(\dfrac{n-d_n}{n}\right)^{n-d_n}  \left(\dfrac{n-r}{n-d_n-r} \right)^{n-d_n-r}	\nonumber\\
	&= \left(1+\dfrac{d_n}{r}\right)^r  \left(1-\dfrac{d_n}{n}\right)^{n-d_n}  \left(1+\dfrac{d_n}{n-d_n-r} \right)^{n-d_n-r} \nonumber\\
	&\leq  \exp\left\{d_n-d_n\left(1-\dfrac{d_n}{n}\right)+d_n\right\}\label{eq:gc-del-5} \\
	&={\exp\left\{d_n\left(1 + \dfrac{d_n}{n}\right)\right\}}, \label{eq:gc-del-binom-simple}
\end{align}
where \eqref{eq:gc-del-4} and \eqref{eq:gc-del-5} follow from  \eqref{eq:cdc18fact-dn}  and \eqref{eq:gc_1pmx} respectively.
Combining \eqref{eq:gc-del-3} and \eqref{eq:gc-del-binom-simple}, we get
\begin{align}
		& {\binom{n-d_n}{r}}\pr[\sdcondn]\nonumber\\ 
		&\leq {\exp\left\{d_n\left(1 + \dfrac{d_n}{n}\right)\right\}} \left( \mu +(1-\mu) \left(1-\dfrac{r}{n}\right)^{\K-1} \right)^{n-d_n}.\label{eq:gc-combined}
\end{align}
%\hrule
%\vspace{10pt}
%\emph{(Choosing the threshold to split the sum)}\\
Recall that Theorem~\ref{theorem:gc-deleted} deals with the case $d_n=o(n)$ and throughout we assume {$x_n > \frac{(1+\epsilon)d_n}{(\kk-1)} ~\forall n$ }and $x_n=o(n)$ , for a fixed $\epsilon>0$. We define $f:\N_0 \rightarrow \N_0$ to be
{
\begin{align}
	f_n = \max \left\{x_n, \frac{(1+\epsilon) d_n}{(1-\mu)}\right\}~\forall n.
	\label{eq:gc-del-fn-def}
\end{align}
Note that under the enforced assumptions $d_n=o(n)$ and $x_n=o(n)$, by definition \eqref{eq:gc-del-fn-def}, $f_n=o(n)$.\\
%vs
%$$ f_n = \frac{2 d_n}{(1-\mu)}~\forall n$$
%vs
%$$ f_n = \frac{(1+\epsilon) d_n}{(1-\mu)}~\forall n, \epsilon >0$$
%$$ f_n = \frac{1-\mu}{\kk-1} d_n$$
}
%{\color{blue}**Discuss**} what if first summation is empty? do we need max with $x_n$,( in that case, \eqref{eq:gc-end})? Can we constraint $x_n$ to be $o(n)$ --need $f_n=o(n)$ in proof of fact 1?
\emph{Remark:} While defining $f_n$ through \eqref{eq:gc-del-fn-def}, we can pick any $\delta>0$ such that $f_n = \max \left\{x_n, \frac{(1+\delta) d_n}{(1-\mu)}\right\}~\forall n$ and the same sequence of steps can be used to complete the proof for Theorem~\ref{theorem:gc-deleted}. For concreteness we choose $\delta=\epsilon$ yielding  \eqref{eq:gc-del-fn-def}.\\
%{\color{blue}remark: $1+\epsilon$}
%\footnote{{\color{blue}Discuss:} Instead of 2, 1+$\epsilon$, $\epsilon > 0$}
\noindent Next, we split the summation (\ref{eq:gc-del-sum}) as follows:
\begin{align}
	&\sum_{r=x_n}^{\floor{(n-d_n)/2}} {\binom{n-d_n}{r}}\pr[\sdcondn]\nonumber\\
	&\quad =  \sum_{r=x_n}^{\floor{f_n}} {\binom{n-d_n}{r}}\pr[\sdcondn]\nonumber \\
	&\quad +\sum_{r=\floor{f_n}+1}^{\floor{(n-d_n)/2}} {\binom{n-d_n}{r}}\pr[\sdcondn].\label{eq:gcsplit-on} \end{align}{}
%where $x_n < f_n < (n-d_n)/2 ~\forall n.$\\
\subsection{ Upper  bound for $x_n \leq r \leq \floor{f_n}$}%, we have%$, if {\color{blue}$f_n=o(n)$} and for some $\epsilon>0$,  {\color{blue}$$x_n > \dfrac{d_n(1+\epsilon)}{(\kk -1 )} $$}
	%{\color{blue}$x_n=\omega(1)$} since $d_n=\omega(1)$
	%then%, (upon combining with \eqref{eq:gc-del-binom-simple})	
%	\begin{align}
%		%	& leq  \left( 1- (1-\mu)\dfrac{r(\K-1)}{n}\left(1-\dfrac{r (\K-1)}{2n} \right) \right)^{n(1-\frac{d_n}{n})}%	
%		&{\binom{n-d_n}{r}}\pr[\sdcondn] \nonumber\\
%	&\leq \exp \left\{ - \frac{r(\kk-1)}{2}\left(1-o(1)\right)\right\}.\label{eq:gc-claim1}
%\end{align} 
%where {\color{blue}$c_n=o(1)$, $0<c_n<1~\forall n$}.}\\
%\emph{Proof of Fact 1:}\\
%To show \eqref{eq:gc-claim1}, 
For $x_n \leq r \leq \floor{f_n}$, we first simplify \eqref{eq:gc-combined}
%Justification for observation 1): (Similar to regime 1 in original proof +  \eqref{eq:gc-del-binom-simple}) Fact (\ref{eq:gc_junfact})
using Fact (\ref{eq:gc_junfact}) with $x=\frac{r}{n}$ to get
\begin{align}
	& \left( \mu +(1-\mu) \left(1-\dfrac{r}{n}\right)^{\K-1} \right)^{n-d_n}\nonumber \\
	&=\left( 1- (1-\mu) \left(1-\left(1-\dfrac{r}{n}\right)^{\K-1}\right) \right)^{n\left(1-\frac{d_n}{n}\right)} \nonumber\\
	&\leq \bigg( 1- (1-\mu) \bigg(1- \bigg(1-\dfrac{r(\K-1)}{n}\nonumber\\&\quad+\dfrac{r^2 (\K-1)^2}{2n^2}  \bigg) \bigg) \bigg)^{n \left(1-\frac{d_n}{n} \right)}  	\label{eq:dl-fact1-01} \\
	%	&= \left( \mu +(1-\mu) \left(1-\dfrac{r}{n}\right)^{\K-1} \right)^{n\left(1-\frac{d_n}{n}\right)} \nonumber \\
	%&\leq \left( 1- (1-\mu)\left(1-\left(1-\dfrac{r(\K-1)}{n}+\dfrac{r^2 (\K-1)^2}{2n^2} \right)\right) \right)^{n\left(1-\frac{d_n}{n}\right)} \label{eq:gc-del-6} \\
	%&= \left( 1- (1-\mu)\left(\dfrac{r(\K-1)}{n}-\dfrac{r^2 (\K-1)^2}{2n^2} \right) \right)^{n}  \nonumber \\
	%	&= \left( 1- (1-\mu)\left(\dfrac{r (\K-1)}{n}- \dfrac{r^2 (\K-1)^2}{2n^2}\right) \right)^{n\left(1-\frac{d_n}{n}\right)} \nonumber\\
	&= \left( 1- (1-\mu)\dfrac{r (\K-1)}{n}\left(1- \dfrac{r (\K-1)}{2n} \right) \right)^{n\left(1-\frac{d_n}{n}\right)}
	\nonumber\\
	&\leq  \exp \left\{- (1-\mu){r (\K-1)}\left(1- \dfrac{r (\K-1)}{2n} \right)\left(1-\frac{d_n}{n}\right) \right\}\nonumber\\
	&=\exp \left\{- {r (\kk-1)}\left(1- \dfrac{r (\K-1)}{2n} \right)\left(1-\frac{d_n}{n}\right) \right\}.	\label{eq:dl-fact1-11}
\end{align}
where \eqref{eq:dl-fact1-01} and \eqref{eq:dl-fact1-11} follow from Fact (\ref{eq:gc_junfact}) and \eqref{eq:gc_1pmx} respectively. Combining \eqref{eq:gc-del-binom-simple} and \eqref{eq:dl-fact1-11}, for $x_n \leq r \leq \floor{f_n}$, we have
\begin{align}
	&{\binom{n-d_n}{r}}\pr[\sdcondn]\nonumber\\
	& \leq  \exp\Bigg\{d_n\left(1 + \dfrac{d_n}{n}\right)\nonumber\\&\quad-{r (\kk-1)}\left(1- \dfrac{r (\K-1)}{2n} \right)\left(1-\frac{d_n}{n}\right) \Bigg\}.	\label{eq:dl-fact1-1}
\end{align}
Using  $\frac{(1+\epsilon)d_n}{(\kk-1)} <x_n \leq r \leq \floor{f_n}$ together with with $f_n = o(n), d_n=o(n)$ and $K_n = O(1)$,  for $x_n \leq r \leq \floor{f_n}$ we have 
%For $r \leq \floor{f_n}$, $r/n <f_n/n = o(1)$ with $f_n = o(n)$ and using $d_n=o(n)$ together with $K_n = O(1)$ we get
\begin{align}
	&{\binom{n-d_n}{r}}\pr[\sdcondn]\nonumber\\
	& \leq  \exp\left\{d_n\left(1 + \dfrac{d_n}{n}\right)-r(\kk-1)\left(1-o(1)\right)\right\}   \nonumber\\
	&= \exp\Bigg\{-(\kk-1)r \cdot \Bigg[1-o(1)-\dfrac{d_n}{(\kk-1)r}\nonumber\\
	&\qquad\qquad \cdot \left(1+\dfrac{d_n}{n}\right) \Bigg]\Bigg\}\nonumber\\%\label{eq:gc-del-10}\nonumber\\
%	&\leq \exp\Bigg\{-(\kk-1)r \cdot \Bigg[1-o(1)-\dfrac{d_n}{(\kk-1)x_n}\nonumber\\
%	&\qquad\qquad \cdot \left(1+\dfrac{d_n}{n}\right) \Bigg]\Bigg\}
%\end{align}
%Thus, for $x_n \leq r \leq \floor{f_n}$ we have 
%\begin{align}
%	&{\binom{n-d_n}{r}}\pr[\sdcondn]\nonumber\\
	&\leq \exp\Bigg\{-(\kk-1)r \cdot \Bigg[1-o(1)-\dfrac{1}{(1+\epsilon)}\nonumber\\
	&\qquad\qquad \cdot \left(1+\dfrac{d_n}{n}\right) \Bigg]\Bigg\}\nonumber\\
	&= \exp \left\{ - r(\kk-1) \frac{\epsilon}{1+\epsilon}\left(1-o(1)\right)\right\}.\label{eq:gc-claim1}
\end{align}
%For $r \leq \floor{f_n}$, $r/n <f_n/n = o(1)$ with $f_n = o(n)$ and using $d_n=o(n)$ together with $K_n = O(1)$ we get
%\begin{align}
%	&{\binom{n-d_n}{r}}\pr[\sdcondn]\nonumber\\
%	& \leq  \exp\left\{d_n\left(1 + \dfrac{d_n}{n}\right)-r(\kk-1)\left(1-o(1)\right)\right\}   \nonumber\\
%	&= \exp\Bigg\{-(\kk-1)r \cdot \Bigg[1-o(1)-\dfrac{d_n}{(\kk-1)r}\nonumber\\
%	&\qquad\qquad \cdot \left(1+\dfrac{d_n}{n}\right) \Bigg]\Bigg\}\nonumber\\%\label{eq:gc-del-10}\nonumber\\
%	&\leq \exp\Bigg\{-(\kk-1)r \cdot \Bigg[1-o(1)-\dfrac{d_n}{(\kk-1)x_n}\nonumber\\
%	&\qquad\qquad \cdot \left(1+\dfrac{d_n}{n}\right) \Bigg]\Bigg\}
%\end{align}
%Now, under the conditions $x_n > \frac{(1+\epsilon)d_n}{(\kk-1)} ~\forall n$ and $d_n=o(n)$, $x_n \leq r \leq \floor{f_n}$ we have 
%\begin{align}
%	&{\binom{n-d_n}{r}}\pr[\sdcondn]\nonumber\\
%	&\leq \exp\Bigg\{-(\kk-1)r \cdot \Bigg[1-o(1)-\dfrac{1}{(1+\epsilon)}\nonumber\\
%	&\qquad\qquad \cdot \left(1+\dfrac{d_n}{n}\right) \Bigg]\Bigg\}\nonumber\\
%	&= \exp \left\{ - r(\kk-1) \frac{\epsilon}{1+\epsilon}\left(1-o(1)\right)\right\}.\label{eq:gc-claim1}
%\end{align}

\subsection{Upper bound for $\floor{f_n}+1 \leq r \leq (n-d_n)/2$}
%\newpage
%c For $\floor{f_n}+1 \leq r \leq (n-d_n)/2$, we have\\
%	%and
%	%{\color{red}$$d_n=o(f_n)$$}bound for
%	\begin{align}
%		%	& leq  \left( 1- (1-\mu)\dfrac{r(\K-1)}{n}\left(1-\dfrac{r (\K-1)}{2n} \right) \right)^{n(1-\frac{d_n}{n})}%	
%		&{\binom{n-d_n}{r}}\pr[\sdcondn] \leq \exp \left\{ - \frac{r(1-\mu)}{2}\left(1-o(1)\right)\right\}.\label{eq:gc-claim2}
%\end{align} 
%\emph{Proof of Fact 2:}
\noindent From \eqref{eq:gc-combined}, for $\floor{f_n}+1 \leq r \leq \floor{(n-d_n)/2}$ we have
\begin{align}
	&{\binom{n-d_n}{r}}\pr[\sdcon]\nonumber\\
	& \leq  \exp\left\{d_n\left(1 + \dfrac{d_n}{n}\right)\right\}   \left( \mu +(1-\mu) \left(1-\dfrac{r}{n}\right)^{\K-1} \right)^{n-d_n} \nonumber\\
	& \leq  \exp\left\{d_n\left(1 + \dfrac{d_n}{n}\right)\right\}   \left( \mu +(1-\mu) \left(1-\dfrac{r}{n}\right) \right)^{n-d_n}\label{eq:gc-del-b-1} \\
	& =  \exp\left\{d_n\left(1 + \dfrac{d_n}{n}\right)\right\}  \left( 1-(1-\mu) \left(\dfrac{r}{n}\right) \right)^{n-d_n} \nonumber\\
	& \leq  \exp\left\{d_n\left(1 + \dfrac{d_n}{n}\right)  -(1-\mu)r \left(1-\dfrac{d_n}{n} \right)\right\} \label{eq:gc-del-b-2} \\
	&=\exp\left\{-(1-\mu)r\left(1-\dfrac{d_n}{n}-\dfrac{d_n}{(1-\mu)r}\left(1 + \dfrac{d_n}{n}\right)\right)\right\} \nonumber\\
	&\leq \exp\Bigg\{-(1-\mu)r\Bigg(1-\dfrac{d_n}{n}-\dfrac{1}{1+\epsilon}\left(1 + \dfrac{d_n}{n}\right) \Bigg)\Bigg\}\nonumber\\
		&= \exp\Bigg\{-r(1-\mu)\dfrac{\epsilon}{1+\epsilon}\Bigg(1-\dfrac{d_n}{n} \Bigg)\Bigg\}\label{eq:gc-jul-1},
			%&\quad \dfrac{d_n(1-\mu)}{(1-\mu) d_n (1+\epsilon)}\left(1 + \dfrac{d_n}{n}\right) \Bigg)\Bigg\}\label{eq:gc-jul-1},
\end{align}
where \eqref{eq:gc-del-b-1} is a consequence of $\K \geq 2~\forall n$, \eqref{eq:gc-del-b-2} follows from \eqref{eq:gc_1pmx} and \eqref{eq:gc-jul-1} is immediate from $r \geq \floor{f_n}+1> \frac{(1+\epsilon)d_n}{(1-\mu )}$. Using $d_n=o(n)$, for $\floor{f_n}+1 \leq r \leq \floor{(n-d_n)/2}$ we have

	\begin{align}
		%	& leq  \left( 1- (1-\mu)\dfrac{r(\K-1)}{n}\left(1-\dfrac{r (\K-1)}{2n} \right) \right)^{n(1-\frac{d_n}{n})}%	
		&{\binom{n-d_n}{r}}\pr[\sdcondn] \nonumber\\
		&\leq \exp\Bigg\{-r(1-\mu)\dfrac{\epsilon}{1+\epsilon}\left(1-o(1) \right)\Bigg\}.\label{eq:gc-claim2}
\end{align}

\subsection{Proof of Proposition~\ref{prop:gc-del}}
Now, using (\ref{eq:gc-claim1}) and \eqref{eq:gc-claim2}, we can upper bound summation (\ref{eq:gcsplit-on}) as follows
\begin{align}
	& \sum_{r=x_n}^{\floor{(n-d_n)/2}} {\binom{n-d_n}{r}}\pr[\sdcon]\nonumber \\
	&=  \sum_{r=x_n}^{\floor{f_n}} {\binom{n-d_n}{r}}\pr[\sdcon] \nonumber \\
	&+ \sum_{r=\floor{f_n}+1}^{\floor{(n-d_n)/2}} {\binom{n-d_n}{r}}\pr[\sdcon] \nonumber\\
	&\leq  \sum_{r=x_n}^{\floor{f_n}} \hspace{-2mm}  e^{-{r(\kk-1)}\frac{\epsilon}{1+\epsilon}\left(1- \oo(1) \right)} + \hspace{-2mm} \sum_{r=\floor{f_n}+1}^{\floor{(n-d_n)/2}} \hspace{-2mm} e^{{-r(1-\mu)}\frac{\epsilon}{1+\epsilon}\left(1-o(1)\right)}\nonumber\\
		&\leq  \sum_{r=x_n}^{\infty} \hspace{-2mm}  e^{-{r(\kk-1)}\frac{\epsilon}{1+\epsilon}\left(1- \oo(1) \right)} + \hspace{-2mm} \sum_{r=\floor{f_n}+1}^{\infty} \hspace{-2mm} e^{{-r(1-\mu)}\frac{\epsilon}{1+\epsilon}\left(1-o(1)\right)}\nonumber
%	&\leq  \sum_{r=x_n}^{\infty} \hspace{-2mm}  e^{-{r(\kk-1)}\frac{\epsilon}{1+\epsilon}\left(1- \oo(1) \right)} + \hspace{-2mm} \sum_{r=\floor{f_n}+1}^{\infty} \hspace{-2mm} e^{r(1-\mu)}\left(1-o(1)\right)}\nonumber
	% & \leq \left(  \sum_{r=M}^{\floor{n/\log n}}  \exp \left\{-{r(\kk-1)}\left(1- \oo(1) \right)\right\}\right) + e^{-(1-\mu)\frac{n}{\log n}}  \nonumber \\
	%  &\leq \left(  \sum_{r=M}^{\infty}  \exp \left\{-{r(\kk-1)}\left(1- \oo(1) \right)\right\}\right) + \oo(1).  \nonumber
\end{align}
 %and the secon summation above vanishes.
Both of the above geometric series have all terms strictly less than one, and thus they are {\em summable}. This gives
	\begin{align}
		&\sum_{r=x_n}^{\floor{(n-d_n)/2}} {\binom{n-d_n}{r}}\pr[\sdcon]\nonumber \\
	%	&\leq \frac{e^{-\frac{x_n}{2}\left(\kk-1\right)(1-\oo(1))}}{1-{e^{-\left(\kk-1\right)(1-\oo(1))}}} + \frac{e^{-\frac{(\floor{f_n}+1)}{2}\left(1-\mu\right)(1-\oo(1))}}{1-{e^{-\left(1-\mu\right)(1-\oo(1))}}}
		\nonumber  \\
	%		&  \pr \left[ |\cmaxdn| \leq n-  x_n\right] \nonumber \\
		&\leq \frac{e^{- x_n (\kk-1) \frac{\epsilon}{1+\epsilon}(1-\oo(1))}}{1-{e^{-\left(\kk-1\right)\frac{\epsilon}{1+\epsilon}(1-\oo(1))}}} + \frac{e^{-{(\floor{f_n}+1)}\left(1-\mu\right)\frac{\epsilon}{1+\epsilon}(1-\oo(1))}}{1-{e^{-\left(1-\mu\right)\frac{\epsilon}{1+\epsilon}(1-\oo(1))}}}.\nonumber\\
				&\leq \frac{e^{- x_n (\kk-1) \frac{\epsilon}{1+\epsilon}(1-\oo(1))}}{1-{e^{-\left(\kk-1\right)\frac{\epsilon}{1+\epsilon}(1-\oo(1))}}} + \frac{e^{-{x_n}\left(1-\mu\right)\frac{\epsilon}{1+\epsilon}(1-\oo(1))}}{1-{e^{-\left(1-\mu\right)\frac{\epsilon}{1+\epsilon}(1-\oo(1))}}}, \label{eq:gc-del-e1}
\end{align}
where \eqref{eq:gc-del-e1} is a consequence of $f_n \geq x_n$ by \eqref{eq:gc-del-fn-def}.\hfill\fsquare

\section{Conclusions}
This work analyzes the existence and size of the giant component for inhomogeneous random K-out graphs. In particular, we prove that whenever $\K\geq2$, there exists a giant component of size $n-O(1)$ whp. Moreover, even when a randomly chosen \emph{finite} subset of nodes is deleted from the network, our results show that a giant component of size $n- \OO(1)$ persists for all $K_n \geq 2$ whp. We also establish the existence of a giant component of size $n(1-o(1))$ whp when $d_n=o(n)$ nodes are randomly deleted from the network. This work goes beyond existing results on connectivity in inhomogeneous random K-out graphs that require $\K=\omega(1)$ by showing that a bounded choice of $\K$ is sufficient to obtain a giant component of size $n-O(1)$. In particular, our results further strengthen the applicability of random K-out graphs as an efficient way to construct a connected network topology in situations where resources are scarce and some nodes may fail or get compromised. An open direction of research is to characterize the asymptotic size of the largest connected component of homogeneous random K-out graph when $K=1$ which is not known to the best of our knowledge. It is also of interest to pursue further applications of random K-out graphs in decentralized learning algorithms and cryptographic payment channel networks.

\bibliographystyle{IEEEtran}
\bibliography{IEEEabrv,references}

\begin{thebibliography}{10}
\providecommand{\url}[1]{#1}
\csname url@samestyle\endcsname
\providecommand{\newblock}{\relax}
\providecommand{\bibinfo}[2]{#2}
\providecommand{\BIBentrySTDinterwordspacing}{\spaceskip=0pt\relax}
\providecommand{\BIBentryALTinterwordstretchfactor}{4}
\providecommand{\BIBentryALTinterwordspacing}{\spaceskip=\fontdimen2\font plus
\BIBentryALTinterwordstretchfactor\fontdimen3\font minus
  \fontdimen4\font\relax}
\providecommand{\BIBforeignlanguage}[2]{{%
\expandafter\ifx\csname l@#1\endcsname\relax
\typeout{** WARNING: IEEEtran.bst: No hyphenation pattern has been}%
\typeout{** loaded for the language `#1'. Using the pattern for}%
\typeout{** the default language instead.}%
\else
\language=\csname l@#1\endcsname
\fi
#2}}
\providecommand{\BIBdecl}{\relax}
\BIBdecl

\bibitem{mansi_cdc}
M.~Sood and O.~Ya\u{g}an, ``On the size of the giant component in inhomogeneous
  random k-out graphs,'' in \emph{2020 59th IEEE Conference on Decision and
  Control (CDC)}, 2020, pp. 5592--5597.

\bibitem{barabasi2016network}
A.-L. Barab{\'a}si, \emph{Network science}.\hskip 1em plus 0.5em minus
  0.4em\relax Cambridge university press, 2016.

\bibitem{newman2002random}
M.~E. Newman, D.~J. Watts, and S.~H. Strogatz, ``Random graph models of social
  networks,'' \emph{Proceedings of the National Academy of Sciences}, vol.~99,
  no. suppl 1, pp. 2566--2572, 2002.

\bibitem{Newman_2002}
M.~E.~J. Newman, ``Spread of epidemic disease on networks,'' \emph{Phys. Rev.
  E}, vol.~66, p. 016128, Jul 2002.

\bibitem{FennerFrieze1982}
T.~I. Fenner and A.~M. Frieze, ``On the connectivity of random $m$-orientable
  graphs and digraphs,'' \emph{Combinatorica}, vol.~2, no.~4, pp. 347--359, Dec
  1982.

\bibitem{Bollobas}
B.~Bollob{\'a}s, \emph{Random graphs}.\hskip 1em plus 0.5em minus 0.4em\relax
  Cambridge university press, 2001, vol.~73.

\bibitem{Yagan2013Pairwise}
O.~Ya\u{g}an and A.~M. Makowski, ``On the connectivity of sensor networks under
  random pairwise key predistribution,'' \emph{IEEE Transactions on Information
  Theory}, vol.~59, no.~9, pp. 5754--5762, Sept 2013.

\bibitem{mansi_icc}
M.~Sood and O.~Ya\u{g}an, ``Tight bounds for the probability of connectivity in
  random k-out graphs,'' in \emph{2021 IEEE International Conference on
  Communications(ICC)}, 2021.

\bibitem{FantiDandelion2018}
G.~Fanti, S.~B. Venkatakrishnan, S.~Bakshi, B.~Denby, S.~Bhargava, A.~Miller,
  and P.~Viswanath, ``Dandelion++: Lightweight cryptocurrency networking with
  formal anonymity guarantees,'' \emph{Proc. ACM Meas. Anal. Comput. Syst.},
  vol.~2, no.~2, pp. 29:1--29:35, Jun. 2018.

\bibitem{2020dprivacy}
C.~Sabater, A.~Bellet, and J.~Ramon, ``Distributed differentially private
  averaging with improved utility and robustness to malicious parties,''
  \emph{arXiv preprint arXiv:2006.07218}, 2020.

\bibitem{yagan2012modeling}
O.~Ya\u{g}an and A.~M. Makowski, ``Modeling the pairwise key predistribution
  scheme in the presence of unreliable links,'' \emph{Information Theory, IEEE
  Transactions on}, vol.~59, no.~3, pp. 1740--1760, March 2013.

\bibitem{yavuz2017k}
F.~Yavuz, J.~Zhao, O.~Ya{\u{g}}an, and V.~Gligor, ``$ k $-connectivity in
  random $ k $-out graphs intersecting erd{\H{o}}s-r{\'e}nyi graphs,''
  \emph{IEEE Transactions on Information Theory}, vol.~63, no.~3, pp.
  1677--1692, 2017.

\bibitem{yavuz2015toward}
------, ``Toward $ k $-connectivity of the random graph induced by a pairwise
  key predistribution scheme with unreliable links,'' \emph{IEEE Transactions
  on Information Theory}, vol.~61, no.~11, pp. 6251--6271, 2015.

\bibitem{Haowen_2003}
H.~Chan, A.~Perrig, and D.~Song, ``Random key predistribution schemes for
  sensor networks,'' in \emph{2003 Symposium on Security and Privacy,
  2003.}\hskip 1em plus 0.5em minus 0.4em\relax IEEE, 2003, pp. 197--213.

\bibitem{Gligor_2002}
\BIBentryALTinterwordspacing
L.~Eschenauer and V.~D. Gligor, ``A key-management scheme for distributed
  sensor networks,'' in \emph{Proceedings of the 9th ACM Conference on Computer
  and Communications Security}, ser. CCS '02.\hskip 1em plus 0.5em minus
  0.4em\relax New York, NY, USA: ACM, 2002, pp. 41--47. [Online]. Available:
  \url{http://doi.acm.org/10.1145/586110.586117}
\BIBentrySTDinterwordspacing

\bibitem{wsnapplications}
S.~R. {Jino Ramson} and D.~J. {Moni}, ``Applications of wireless sensor
  networks — a survey,'' in \emph{2017 International Conference on
  Innovations in Electrical, Electronics, Instrumentation and Media Technology
  (ICEEIMT)}, 2017, pp. 325--329.

\bibitem{federated2019advances}
P.~Kairouz, H.~B. McMahan, B.~Avent, A.~Bellet, M.~Bennis, A.~N. Bhagoji,
  K.~Bonawitz, Z.~Charles, G.~Cormode, R.~Cummings \emph{et~al.}, ``Advances
  and open problems in federated learning,'' \emph{arXiv preprint
  arXiv:1912.04977}, 2019.

\bibitem{eletrebycdc2018}
R.~{Eletreby} and O.~{Yağan}, ``Connectivity of wireless sensor networks
  secured by the heterogeneous random pairwise key predistribution scheme,'' in
  \emph{2018 IEEE Conference on Decision and Control (CDC)}, 2018, pp.
  2085--2092.

\bibitem{eletrebyIT20}
R.~Eletreby and O.~Ya\u{g}an, ``Connectivity of inhomogeneous random k-out
  graphs,'' \emph{IEEE Transactions on Information Theory}, vol.~66, no.~11,
  pp. 7067--7080, 2020.

\bibitem{du2007effective}
X.~Du, Y.~Xiao, M.~Guizani, and H.-H. Chen, ``An effective key management
  scheme for heterogeneous sensor networks,'' \emph{Ad Hoc Networks}, vol.~5,
  no.~1, pp. 24--34, 2007.

\bibitem{8606999}
R.~Eletreby and O.~Ya\u{g}an, ``$k$-connectivity of inhomogeneous random key
  graphs with unreliable links,'' \emph{IEEE Transactions on Information
  Theory}, vol.~65, no.~6, pp. 3922--3949, June 2019.

\bibitem{Rashad/Inhomo}
------, ``Connectivity of wireless sensor networks secured by heterogeneous key
  predistribution under an on/off channel model,'' \emph{IEEE Transactions on
  Control of Network Systems}, 2018.

\bibitem{Yagan/Inhomogeneous}
O.~Ya\u{g}an, ``Zero-one laws for connectivity in inhomogeneous random key
  graphs,'' \emph{IEEE Transactions on Information Theory}, vol.~62, no.~8, pp.
  4559--4574, Aug 2016.

\bibitem{Barabasi_1999}
A.-L. Barab{\'a}si and R.~Albert, ``Emergence of scaling in random networks,''
  \emph{Science}, vol. 286, no. 5439, pp. 509--512, 1999.

\bibitem{boccaletti2006complex}
S.~Boccaletti, V.~Latora, Y.~Moreno, M.~Chavez, and D.-U. Hwang, ``Complex
  networks: Structure and dynamics,'' \emph{Physics reports}, vol. 424, no.
  4-5, pp. 175--308, 2006.

\bibitem{Lu2008_applications}
K.~Lu, Y.~Qian, M.~Guizani, and H.-H. Chen, ``A framework for a distributed key
  management scheme in heterogeneous wireless sensor networks,'' \emph{IEEE
  Transactions on Wireless Communications}, vol.~7, no.~2, pp. 639--647,
  February 2008.

\bibitem{Wu2007_applications}
C.-H. Wu and Y.-C. Chung, ``Heterogeneous wireless sensor network deployment
  and topology control based on irregular sensor model,'' in \emph{Advances in
  Grid and Pervasive Computing}, 2007, pp. 78--88.

\bibitem{Yarvis_2005}
M.~Yarvis, N.~Kushalnagar, H.~Singh, A.~Rangarajan, Y.~Liu, and S.~Singh,
  ``Exploiting heterogeneity in sensor networks,'' in \emph{Proceedings IEEE
  24th Annual Joint Conference of the IEEE Computer and Communications
  Societies.}, vol.~2, March 2005, pp. 878--890 vol. 2.

\bibitem{mao2017connectivity}
G.~Mao, \emph{Connectivity of communication networks}.\hskip 1em plus 0.5em
  minus 0.4em\relax Springer, 2017.

\bibitem{hwang2004revisiting}
J.~Hwang and Y.~Kim, ``Revisiting random key pre-distribution schemes for
  wireless sensor networks,'' in \emph{Proceedings of the 2nd ACM workshop on
  Security of ad hoc and sensor networks}.\hskip 1em plus 0.5em minus
  0.4em\relax ACM, 2004, pp. 43--52.

\bibitem{MeiPanconesiRadhakrishnan2008}
A.~Mei, A.~Panconesi, and J.~Radhakrishnan, ``Unassailable sensor networks,''
  in \emph{Proc. of the 4th International Conference on Security and Privacy in
  Communication Netowrks}, ser. SecureComm '08.\hskip 1em plus 0.5em minus
  0.4em\relax New York, NY, USA: ACM, 2008.

\bibitem{liu2006coverage}
X.~Liu, ``Coverage with connectivity in wireless sensor networks,'' in
  \emph{2006 3rd International Conference on Broadband Communications, Networks
  and Systems}.\hskip 1em plus 0.5em minus 0.4em\relax IEEE, 2006, pp. 1--8.

\bibitem{erdHos1960evolution}
P.~Erd{\H{o}}s and A.~R{\'e}nyi, ``On the evolution of random graphs,''
  \emph{Publ. Math. Inst. Hung. Acad. Sci}, vol.~5, no.~1, pp. 17--60, 1960.

\bibitem{elumar2021connectivity}
E.~C. Elumar, M.~Sood, and O.~Ya\u{g}an, ``On the connectivity and giant
  component size of random k-out graphs under randomly deleted nodes,''
  \emph{arXiv preprint arXiv:2103.01471}, 2021.

\bibitem{IT21sood}
M.~Sood and O.~Yağan, ``On the minimum node degree and k-connectivity in
  inhomogeneous random k-out graphs,'' \emph{IEEE Transactions on Information
  Theory}, pp. 1--1, 2021.

\bibitem{2018dprivacy}
P.~Dellenbach, A.~Bellet, and J.~Ramon, ``Hiding in the crowd: A massively
  distributed algorithm for private averaging with malicious adversaries,''
  \emph{arXiv preprint arXiv:1803.09984}, 2018.

\bibitem{SCION2017book}
A.~Perrig, P.~Szalachowski, R.~M. Reischuk, and L.~Chuat, \emph{SCION: a secure
  Internet architecture}.\hskip 1em plus 0.5em minus 0.4em\relax Springer,
  2017.

\bibitem{SIAM2021paper}
S.~Sridhara, F.~Wirz, J.~de~Ruiter, C.~Schutijser, M.~Legner, and A.~Perrig,
  ``Global distributed secure mapping of network addresses,'' in
  \emph{Proceedings of the ACM SIGCOMM Workshop on Technologies, Applications,
  and Uses of a Responsible Internet (TAURIN)}, 2021.

\bibitem{lightning2016}
J.~Poon and T.~Dryja, ``The bitcoin lightning network: Scalable off-chain
  instant payments,'' 2016.

\bibitem{bondy1976graph}
J.~A. Bondy, U.~S.~R. Murty \emph{et~al.}, \emph{Graph theory with
  applications}.\hskip 1em plus 0.5em minus 0.4em\relax Macmillan London, 1976,
  vol. 290.

\bibitem{van2016random}
R.~Van Der~Hofstad, \emph{Random graphs and complex networks}.\hskip 1em plus
  0.5em minus 0.4em\relax Cambridge university press, 2016, vol.~1.

\bibitem{Rybarczyk_2011}
K.~Rybarczyk, ``Diameter, connectivity, and phase transition of the uniform
  random intersection graph,'' \emph{Discrete Mathematics}, vol. 311, no.~17,
  pp. 1998--2019, 2011.

\bibitem{janson2011probability}
S.~Janson, ``Probability asymptotics: notes on notation,'' \emph{arXiv preprint
  arXiv:1108.3924}, 2011.

\bibitem{ZhaoYaganGligor}
J.~Zhao, O.~Ya\u{g}an, and V.~Gligor, ``$k$-connectivity in random key graphs
  with unreliable links,'' \emph{IEEE Transactions on Information Theory},
  vol.~61, no.~7, pp. 3810--3836, July 2015.

\bibitem{coupling2017}
K.~Rybarczyk, ``Sharp threshold functions for random intersection graphs via a
  coupling method,'' \emph{The Electronic Journal of Combinatorics}, vol.~18,
  no.~1, p.~36, 2011.

\bibitem{zhaoconnectivity}
J.~Zhao, O.~Ya\u{g}an, and V.~Gligor, ``On connectivity and robustness in
  random intersection graphs,'' \emph{IEEE Transactions on Automatic Control},
  vol.~62, no.~5, pp. 2121--2136, May 2017.

\end{thebibliography}

%\appendices
\section*{Appendix}
\subsection{Inhomogeneous random K-out graph with $r$ classes}%]
\label{sec:appendix-r-classes}
We now discuss more general constructions of the inhomogeneous random K-out graph with an arbitrary number of node types, denoted by $r$ \cite{eletrebyIT20}. We let each node belong to type-$i$ independently with probability $\mu_i>0$ for $i=1,\ldots,r$ with $\sum_{i=1}^r \mu_i=1$. Each type-$i$ nodes gets paired with $K_{i,n}$ other nodes, chosen uniformly at random from among all other nodes where $1 \leq K_{1,n} < K_{2,n} < \ldots < K_{r,n}$. An undirected edge is drawn between a pair of nodes if either of them chooses the other. We let the resulting inhomogeneous random graph with $r$ classes be denoted by $\HH$ where $\pmb{K_n^r}=[K_{1,n}, K_{2,n},\ldots, K_{r,n}]$ and $\pmb{\mu^r}=[ \mu_1,\mu_2, \ldots, \mu_r ]$ with $\mu_i>0$ for $i=1,\ldots,r$. Let $\HHcmax$ denote the largest connected component of $\HH$.
\begin{corollary}
	{\sl  
		If $K_{r,n} \geq 2 \ \forall n$ then for the inhomogeneous random K-out graph $\HH$ with $r$ node types, we have
		\begin{align}
			|\HHcmax|
			= n-\OO(1) \ \ {\rm whp.}  \nonumber
		\end{align}
	}\label{cor:gc_r_classes}
\end{corollary}{}
For a given $r$, throughout, we let $\hhgc$ denote the inhomogeneous random graph with $2$ node types and parameters $\tilde{\mu}=\sum_{i=1}^{r-1}\mu_i$ and $K_{n} = K_{r,n}$. For proving Corollary~\ref{cor:gc_r_classes}, we use the following proposition.
\begin{proposition}
Let $\pmb{\K}=[K_{1,n}, K_{2,n},\ldots, K_{r,n}]$ and $\pmb{\mu}=[ \mu_1,\mu_2, \ldots, \mu_r ]$ with $\mu_i>0$ for $i=1,\dots,r$.
{\sl For any monotone-increasing property $\mathcal{P}$, i.e., a property which holds upon addition of edges to the graph (see \cite[p.~13]{coupling2017},\cite{zhaoconnectivity}), we have
\begin{align}
	&\pr[ \HH \textrm{ has property } \mathcal{P} ]  \nonumber\\
	&\geq \pr[\hhgc \textrm{ has property } \mathcal{P}] \label{eq:gcmonotoneproperty},
\end{align}
\label{prop:coupling}}
\end{proposition}
where $\tilde{\mu}=\sum_{i=1}^{r-1}\mu_i$.
%{\color{blue} $\mu=, \K=$}

\emph{Proof of Proposition~\ref{prop:coupling}}:
 %complete the proof.  Thus, in order  to establish this Corollary, it suffices to establish the aforementioned coupling. 
 The proof relies on proving the existence of a {\em coupling} under which 
 $\hhgc$ is a spanning subgraph of $\HH$; i.e., if we can generate an instantiation of  $\HH$ by adding edges to an instantiation of $\hhgc$. Define $\tilde{\mu}=\sum_{i=1}^{r-1}\mu_i$.  Consider an instantiation of an inhomogeneous random graph $\hhgc$ with two classes such that each of the $n$ nodes is independently assigned as type-1 (respectively, type-2) with probability $\tilde{\mu}$ (respectively, $1-\tilde{\mu}$) and then type-1 (respectively, type-2) nodes draw edges to $1$ (respectively, $K_{r,n}$) nodes chosen uniformly at random. From this instantiation, we can generate an instantiation of $\HH$ as follows. First, let each type-1 node be independently reassigned as type-$i$ with probability $\frac{\mu_i}{\tilde{\mu}}$ for $i=1,2,\dots,r-1$. Next, for $i=2,\dots,r-1$, let each type-$i$ node pick $K_{i,n}-1$ additional neighbors that were not chosen by it initially. %After these additional choices are made, 
Next, we draw an undirected edge between each pair of nodes where at least one picked the other. Clearly, this process creates a graph whose edge set is a superset of the edge set of the  realization of $\hhgc$ that we started with. In addition, in the new graph, the probability of a node picking $K_{i,n}$ other nodes (i.e., being type-$i$) is given by  $\tilde{\mu} \frac{\mu_i}{\tilde{\mu}} = \mu_i$, for $i=1,2,\dots,r$. We thus conclude that the new graph obtained constitutes a realization of $\HH$. Since, the initial realization of  $\hhgc$ was arbitrary, this establishes the desired coupling between the graphs $\hh$ and $\HH$ as the edge set of $\hh$ is contained in the edge set of $\HH$. Therefore  (\ref{eq:gcmonotoneproperty}) holds for any property that is monotone increasing upon edge addition.\hfill\fsquare \\

\emph{Proof of Corollary~\ref{cor:gc_r_classes}}:
The proof of Corollary~\ref{cor:gc_r_classes} uses Proposition~\ref{prop:coupling} applied to the property $ |\cmax| \geq n - x_n$ which is monotone increasing upon edge addition since adding more edges to the network cannot decrease the size of the giant component. Combining (\ref{eq:gcmonotoneproperty}) with \eqref{eq:cor-initial-proof}, we get for any $x_n=\omega(1)$,
\begin{align*}
&\limit  \pr \left[ |\HHcmax|  \geq  n-x_n\right]\nonumber\\
& \geq \limit\pr \left[   |\cmax|  \geq  n-x_n\right]=1,
\end{align*}
or equivalently,
\begin{align}
	n -  |\HHcmax| \leq x_n \ {\rm whp~for~any} \ \ x_n =\omega(1), \nonumber,\\
\end{align}
and thus $|\HHcmax|
= n-\OO(1)$ whp.
 \hfill\fsquare \\ %Therefore, if there exists a {\em coupling} under which 
%$\hh$ is a spanning subgraph of $\HH$; i.e., if we can generate an instantiation of  $\HH$ by  adding edges to an instantiation of $\hh$, then we can use (\ref{eq:gcmonotoneproperty}) to establish this Corollary. and we conclude 
%that (\ref{eq:gcmonotoneproperty}) holds for the property $ |\cmax| \geq n - M$.

%{\color{blue}If space permits}
Using Proposition~\ref{prop:coupling} and Theorem~\ref{theorem:gc}, we can also derive an upper bound on the probability that more than $M$ nodes are outside of the largest component for $\HH$.

\begin{corollary}
	{\sl 
		For the inhomogeneous random graph $\HH$ with $r$ node types, if $K_{r,n} \geq 2 \ \forall n$ and $K_{r,n} =O(1)$ then for each $M = 1, 2, \ldots$, it holds that 
		\begin{align}
			&  \pr \left[|\HHcmax|\leq n-  M\right] \nonumber \\
			&\leq \frac{\exp\{-M (K_{r,n}-1)(\mu_r)(1-\oo(1))\}}{1-{\exp\{-(K_{r,n}-1)(\mu_r)(1-\oo(1))\}}} + \oo(1).\label{eq:gc-r-classses-bound}
		\end{align}
	}  \label{cor:gc-r-classses-bound}
\end{corollary}

\emph{Proof of Corollary~\ref{cor:gc-r-classses-bound}}:
The proof of Corollary~\ref{cor:gc-r-classses-bound} uses the coupling argument described in the proof of Proposition~\ref{prop:coupling}. For a given $r$, $\HH$ denotes the inhomogeneous random graph with $\pmb{K_n}=[K_{1,n}, K_{2,n},\ldots, K_{r,n}]$ and $\pmb{\mu}=[ \mu_1,\mu_2, \ldots, \mu_r ]$. Consider the inhomogeneous random graph $\hhgc$ with $2$ node types and parameters $\tilde{\mu}=\sum_{i=1}^{r-1}\mu_i$ and $K_{n} = K_{r,n}$ and let $C_{\rm max}(n;\tilde{\mu}, {K}_{r,n} )$ denote its largest connected component. Using (\ref{eq:gcmonotoneproperty}) for the property that the largest connected component is of size greater than $n-M$  together with \eqref{eq:gc_theorem}, and using $\kk-1 = (K_n-1)(1-\mu)$, we have
%In order to prove~\eqref{eq:gc-r-classses-bound}, we can equivalently prove
		\begin{align}
	&  \pr \left[|\HHcmax|> n-  M\right] \nonumber \\
		& \geq \pr \left[|C_{\rm max}(n;\tilde{\mu}, {K}_{r,n} )|> n-  M\right] \nonumber \\
	&\geq 1- \frac{\exp\{-M (K_{r,n}-1)(1-\tilde{\mu})(1-\oo(1))\}}{1-{\exp\{-(K_{r,n}-1)(1-\tilde{\mu})(1-\oo(1))\}}} - \oo(1)\nonumber\\
		&= 1- \frac{\exp\{-M (K_{r,n}-1)(\mu_r)(1-\oo(1))\}}{1-{\exp\{-(K_{r,n}-1)(\mu_r)(1-\oo(1))\}}} - \oo(1)\label{eq:gc-r-classses-bound-2},
\end{align}
and subtracting both sides of \eqref{eq:gc-r-classses-bound-2} from 1 yields \eqref{eq:gc-r-classses-bound}.\hfill\fsquare\\
%		\begin{align}
%	& \pr \left[|\HHcmax|\leq n-  M\right] \nonumber \\
%	&\leq  \frac{\exp\{-M (K_{r,n}-1)(\mu_r)(1-\oo(1))\}}{1-{\exp\{-(K_{r,n}-1)(\mu_r)(1-\oo(1))\}}} + \oo(1).\nonumber
%\end{align}
Using the above argument, we can also characterize the size of the giant component of $\HHcmax$ under random node deletions. Let $\mathcal{P^*}$ denote the property that after a subset of the node set $\nodes$ of size $d_n$ is selected uniformly at random and deleted, the largest connected component in the resulting graph, denoted by $\HHcmaxdn$, is of size greater than or equal to $n-x_n$. It is easy to see that $\mathcal{P^*}$, as defined above, is a monotone-increasing property in edge addition and therefore we can invoke Proposition~\ref{prop:coupling} together with Corollary~\ref{cor:gc-deleted} to show that with $d_n=o(n)$, if $K_{r,n} \geq 2 \ \forall n$ and $K_{r,n} =O(1)$, 
	\begin{enumerate}
	\item[i)] if $d_n=O(1)$, then \begin{align}|\HHcmaxdn| = n- O(1) \ {\rm whp}, \end{align}
	\item[ii)] and if $d_n= \omega(1)$ and $d_n=o(n)$, then \begin{align}|\HHcmaxdn| = n(1-o(1)) \ {\rm whp}.r\end{align}
\end{enumerate}
Next, following the arguments used for proving Corollary~\ref{cor:gc-r-classses-bound}, together with Theorem~\ref{theorem:gc-deleted}, we get for $d_n=o(n)$, if $K_{r,n} \geq 2 \ \forall n$ and $K_{r,n} =O(1)$, we have
\begin{align}
	&  \pr \left[ |\HHcmaxdn| \leq n-  d_n-x_n\right] \nonumber \\
	&\leq \frac{e^{- x_n (K_{r,n}-1)(\mu_r) \frac{\epsilon}{1+\epsilon}(1-\oo(1))}}{1-{e^{-(K_{r,n}-1)(\mu_r)\frac{\epsilon}{1+\epsilon}(1-\oo(1))}}} + \frac{e^{-{x_n}\left(\mu_{r}\right)\frac{\epsilon}{1+\epsilon}(1-\oo(1))}}{1-{e^{-\left(\mu_r\right)\frac{\epsilon}{1+\epsilon}(1-\oo(1))}}}.	
	%	\frac{\exp\{-x_n\left(\kk-1\right)(1-\oo(1))\}}{1-{\exp\{-\left(\kk-1\right)(1-\oo(1))\}}} + \oo(1). 
\end{align}
%{\color{blue}second corollary/node deletion result - repeat proof or through coupling?}\\

%remark on improving the bounds- trade-off xn smaller of bound decay faster?\\
%if space permits, bounds for $r$ classes for Thm3-2, say it can be proved on similar lines for the other case.}

\subsection{Mean node degree in $\hh$}
\label{sec:mean-node-degree}
%Let $\kk$ denote the mean number of edges that each node chooses to draw. Conditioning on the class of node~$i$, we get
% \begin{align}
% \kk&=\mu+(1-\mu)K_n. 
% \end{align}
The probability that node~$v_i$ picks node~$v_j$  where $v_i,v_j\in \nodes$ depends on the type of node~$v_i$ and is given by
\begin{align}
	\pr[v_j \in \Gamma_{n}(v_i)]=\mu\dfrac{1}{n-1}+(1-\mu)\dfrac{K_n}{n-1}=\dfrac{\kk}{n-1}.%=:\kk.
\end{align}

%Recall that each node draws edges to other nodes independently of other nodes. 
Let $v_i \sim v_j$ denote the event that node~$v_i$ is adjacent to node~$j$ in $\hh$.% For $i \sim j$ to occur, either node $i$ selects node $j$ or node $i$ selects node $j$ or both select each other. This gives
\begin{align}
	\pr[i \sim j]&= 1 - (1-\pr[v_i \in \Gamma_{n}(v_j)])(1-\pr[j \in \Gamma_{n}(v_i)]),\nonumber \\
	&=1-\left(1-\dfrac{\kk}{n-1}\right)^2\nonumber\\
	&=\dfrac{2\kk}{n-1}-\left(\dfrac{\kk}{n-1}\right)^2.
\end{align}
Consequently, the mean degree of node~$i$, when $\K=\oo(n)$ is $(1-\oo(1)) 2 \kk$.
%\newpage
%\subsection{Upper bound on $\pr[\cmaxdn \leq n-d_n-x_n]$ }
%\emph{version with $d_n$}
%\emph{for finite $n$}\\

\subsection{Upper bound on $\pr[\cmaxdn \leq n-d_n-x_n]$  with $x_n >\frac{(1+\epsilon)d_n}{(1-\mu)} ~\forall n$}
Observe that $\kk-1 = (\K-1)(1-\mu) \geq (1-\mu) \forall \K \geq 2$. In this discussion, we derive a sharper upper bound than \eqref{eq:gc_general_theorem}, but it requires the stronger condition of $x_n >\frac{(1+\epsilon)d_n}{(1-\mu)}$ as opposed to $x_n >\frac{(1+\epsilon)d_n}{(\kk-1)}$. Note that here we have $\frac{(1+\epsilon)d_n}{(\kk-1)}<x_n \leq r$ and therefore following the sequence of steps from \eqref{eq:gc-del-b-1} to \eqref{eq:gc-jul-1}, for $d_n=o(n)$ and $x_n \leq r \leq (n-d_n)/2$, we get
\begin{align}
	%	& leq  \left( 1- (1-\mu)\dfrac{r(\K-1)}{n}\left(1-\dfrac{r (\K-1)}{2n} \right) \right)^{n(1-\frac{d_n}{n})}%	
	&{\binom{n-d_n}{r}}\pr[\sdcondn] \nonumber\\
	&\leq \exp\Bigg\{-r(1-\mu)\dfrac{\epsilon}{1+\epsilon}\left(1-o(1) \right)\Bigg\}. \nonumber
\end{align} 
Thus, we can upper bound the summation \eqref{eq:gc-del-sum} as follows.
	\begin{align}
	&\sum_{r=x_n}^{\floor{(n-d_n)/2}} {\binom{n-d_n}{r}}\pr[\sdcon]\nonumber \\
	%	&\leq \frac{e^{-\frac{x_n}{2}\left(\kk-1\right)(1-\oo(1))}}{1-{e^{-\left(\kk-1\right)(1-\oo(1))}}} + \frac{e^{-\frac{(\floor{f_n}+1)}{2}\left(1-\mu\right)(1-\oo(1))}}{1-{e^{-\left(1-\mu\right)(1-\oo(1))}}}
	%		&  \pr \left[ |\cmaxdn| \leq n-  x_n\right] \nonumber \\
	&\leq \sum_{r=x_n}^{\floor{(n-d_n)/2}}  e^{{-r(1-\mu)}\frac{\epsilon}{1+\epsilon}\left(1-o(1)\right)}\nonumber\nonumber\\
		&\leq \sum_{r=x_n}^{\infty}  e^{{-r(1-\mu)}\frac{\epsilon}{1+\epsilon}\left(1-o(1)\right)}\nonumber\nonumber\\
	& = \frac{e^{-{x_n}\left(1-\mu\right)\frac{\epsilon}{1+\epsilon}(1-\oo(1))}}{1-{e^{-\left(1-\mu\right)\frac{\epsilon}{1+\epsilon}(1-\oo(1))}}}. \label{eq:gc-del-app-1}
\end{align}
Note that with the condition $x_n >\frac{(1+\epsilon)d_n}{(1-\mu)}$, we do not benefit from splitting the summation at the threshold $f_n$ as defined in \eqref{eq:gc-del-fn-def}. Combining \eqref{eq:gc-del-app-1} with \eqref{eq:gc-del-lem-implication} we get 	\begin{align}
	&  \pr[|\cmaxdn| \leq n -d_n -x_n] \nonumber\\%\pr\left[\left(\zdn \right)\comp\right]\nonumber \\
	%	&\leq  \frac{e^{-\frac{x_n}{2}\left(\kk-1\right)(1-\oo(1))}}{1-{e^{-\left(\kk-1\right)(1-\oo(1))}}} + \frac{e^{-\frac{x_n}{2}\left(1-\mu\right)(1-\oo(1))}}{1-{e^{-\left(1-\mu\right)(1-\oo(1))}}}. \nonumber		
	&\leq \frac{e^{-{x_n}\left(1-\mu\right)\frac{\epsilon}{1+\epsilon}(1-\oo(1))}}{1-{e^{-\left(1-\mu\right)\frac{\epsilon}{1+\epsilon}(1-\oo(1))}}}. \label{eq:gc-del-app-2}
\end{align}{}
Although, \eqref{eq:gc-del-app-2} presents an improvement over the upper bound on $ \pr[|\cmaxdn| \leq n -d_n -x_n]$ presented in Theorem~\ref{theorem:gc-deleted}, it comes at a cost, namely, requiring a larger $x_n$ $\left(x_n >\frac{(1+\epsilon)d_n}{(1-\mu)}\right)$ for the summation to decay to 0. We present the version of the bound with the condition $\left(x_n >\frac{(1+\epsilon)d_n}{(\kk -1)}\right)$ since it captures the dependence on both $\K$ annd $\mu$, requires a smaller value for $x_n$ as compared to \eqref{eq:gc-del-app-2} and yet offers an upper bound on $\pr[|\cmaxdn| \leq n -d_n -x_n]$ that decays to 0 exponentially fast with $x_n$.

\end{document}